\documentclass[twocolumn, aps, pra]{revtex4-1}
\usepackage{amsmath}
\usepackage{amssymb}
\usepackage{braket}
\usepackage{graphicx}
\usepackage{color}
\usepackage{ulem}
\usepackage{hyperref}

\begin{document}
	
\title{Effects of Non-Hermiticity on Su-Schrieffer-Heeger Defect States}

\author{Li-Jun Lang}
\email{langlijun84@gmail.com}
\affiliation{Division of Physics and Applied Physics, School of Physical and Mathematical Sciences, Nanyang Technological University, Singapore 637371, Singapore}
\author{You Wang}
\affiliation{Division of Physics and Applied Physics, School of Physical and Mathematical Sciences, Nanyang Technological University, Singapore 637371, Singapore}
\author{Hailong Wang}
\affiliation{Division of Physics and Applied Physics, School of Physical and Mathematical Sciences, Nanyang Technological University, Singapore 637371, Singapore}
\author{Y. D. Chong}
\email{yidong@ntu.edu.sg}
\affiliation{Division of Physics and Applied Physics, School of Physical and Mathematical Sciences, Nanyang Technological University, Singapore 637371, Singapore}

\date{\today}

\begin{abstract}
  We study the emergence and disappearance of defect states in the complex Su-Schrieffer-Heeger (cSSH) model, a non-Hermitian one-dimensional lattice model containing gain and loss on alternating sites.  Previous studies of this model have focused on the existence of a non-Hermitian defect state that is localized to the interface between two cSSH domains, and is continuable to the topologically protected defect state of the Hermitian Su-Schrieffer-Heeger (SSH) model.  For large gain/loss magnitudes, we find that these defect states can disappear into the continuum, or undergo pairwise spontaneous breaking of a composite sublattice/time-reversal symmetry.  The symmetry-breaking transition gives rise to a pair of defect states continuable to non-topologically-protected defect states of the SSH model.  We discuss the phase diagram for the defect states, and its implications for non-Hermitian defect states.

\end{abstract}

\maketitle

\section{Introduction}

The complex Su-Schrieffer-Heeger (cSSH) model \cite{171007-1} is a non-Hermitian extension of the Su-Schrieffer-Heeger (SSH) model \cite{130704-1}, the simplest one-dimensional (1D) Hermitian lattice exhibiting topological defect states \cite{111222-1}.  It has been the subject of recent interest in experiments \cite{171212-2,171207-7,180409-2,180419-3,180402-1} as a simple testing-ground for the interaction of topological states with \textit{non-Hermiticity}---i.e., the presence of loss and/or gain in the underlying medium \cite{101224-1,110707-1}.  The standard topological invariants used to characterize topological states of matter \cite{120316-1,111222-1}, including the SSH model, assume Hermiticity; for instance, Hermiticity guarantees the existence of a well-defined inner product, which is used to calculate the Zak phases \cite{130107-1} for characterizing the SSH model.  Non-Hermitian generalizations of topological concepts, such as the bulk-edge correspondence principle, are thus of significant theoretical interest \cite{180720-9,171012-1,180720-8,180720-7,180720-3,180720-4,180720-5,180720-6,180720-2}.  Moreover, non-Hermitian variants of topological states may have applications in photonics, where topological protection can be implemented by lattice engineering \cite{tpreview2014, tpreview2017, tpreview2018, hailong1, hailong2}, and non-Hermiticity can be introduced by introducing optical loss and/or gain to the optical medium \cite{180501-3,180501-1}.  The robustness of topological modes may be usefully exploited in amplifiers \cite{liang2013,Peano2016}, lasers \cite{171007-1, 180719-1,180719-2, 180719-2, 180419-3, 180402-1,180720-9, 180720-10}, and other non-Hermitian photonic devices.

Previous studies of the cSSH model, starting with the work of Schomerus \cite{171007-1}, have focused on the existence of a defect state that is exponentially localized to an interface between different cSSH domains.  In the Hermitian limit (no gain or loss), this defect state is explicitly continuable to the well-known topological mid-gap defect state of the SSH model \cite{130704-1}.  In the non-Hermitian case (gain and loss on alternating lattice sites), the energy of the defect state can acquire a nonzero imaginary part, but the real part remains pinned to the mid-gap value.  If the defect configuration is chosen appropriately, the defect state can have a larger amplification rate than any of the bulk states \cite{171007-1}; this has been demonstrated using microwave resonators \cite{171212-2} and lasers \cite{180402-1, 180419-3}.  Alternatively, if the gain and loss are distributed in a parity/time-reversal ($\mathcal{PT}$) symmetric pattern \cite{101224-1}, the bulk and defect state energies can be purely real \cite{171207-7, 180409-2}.

These studies did not, however, look into whether SSH defect states always have a counterpart in the cSSH model, and, conversely, whether the defect states of the cSSH model are always SSH-like.  This is a noteworthy omission because non-Hermitian states are known to be able to exhibit behaviors that have no Hermitian analogue.  $\mathcal{PT}$-symmetric dimer eigenstates, for example, can exhibit spontaneous $\mathcal{PT}$ symmetry breaking \cite{180501-2}, while some non-Hermitian lattices have been shown to support defect states that seem to be topological but have no evident Hermitian counterpart \cite{171009-1, 180306-1, 180322-1, 180613-1, 180720-8}.

In this paper, we analyze the effect of non-Hermiticity on cSSH defect states.  We find that two interesting things can happen to the SSH-like defect state as the gain and loss magnitude is increased.  First, the defect state can disappear via a divergence in its localization length, which corresponds to the merging of the defect state energy into the complex continuum of bulk energies.  Second, the SSH-like defect state can interact with a second defect state that emerges from the continuum.  Both of these states satisfy a composite sublattice/time-reversal ($\mathcal{ST}$) symmetry, which pins the real parts of their energies to zero, and is the non-Hermitian counterpart of the $\mathcal{S}$ symmetry that pins the energy of the SSH mid-gap defect state to zero.  The two states can coalesce in a spontaneous $\mathcal{ST}$-breaking transition (an exceptional point \cite{110707-1}), breaking apart into two $\mathcal{ST}$-broken defect states.  The latter are continuable to the \textit{non}-topologically-protected defect states of the SSH model, which have hitherto been ignored but can exist as well in the cSSH model.  Both methods of destabilizing the SSH-like ``mid-gap'' defect state require $\mathcal{PT}$ symmetry to be spontaneously broken in the bulk bandstructure.

We thus find that the cSSH chain has two anomalous defect state phases not present in the SSH model: (i) a phase with two ``mid-gap'' states localized to the domain wall, rather than one, and (ii) a phase with two ``non-mid-gap'' states but no ``mid-gap'' state.  Phase (ii) includes the special case where the inter-site couplings are uniform (so that the cSSH lattice reduces to a lattice of gain/loss dimers with a defect in the gain/loss pattern); in this limit, the two ``non-mid-gap'' states appear abruptly when the gain/loss magnitude is increased above a certain nonzero threshold, similar to the ``intrinsically non-Hermitian'' defect states that have previously been seen in other non-Hermitian lattice models \cite{171009-1, 180306-1,180613-1, 180720-8}.

\section{The complex Su-Schrieffer-Heeger (cSSH) model} \label{bulk}

\begin{figure}
  \includegraphics[width=0.98\linewidth]{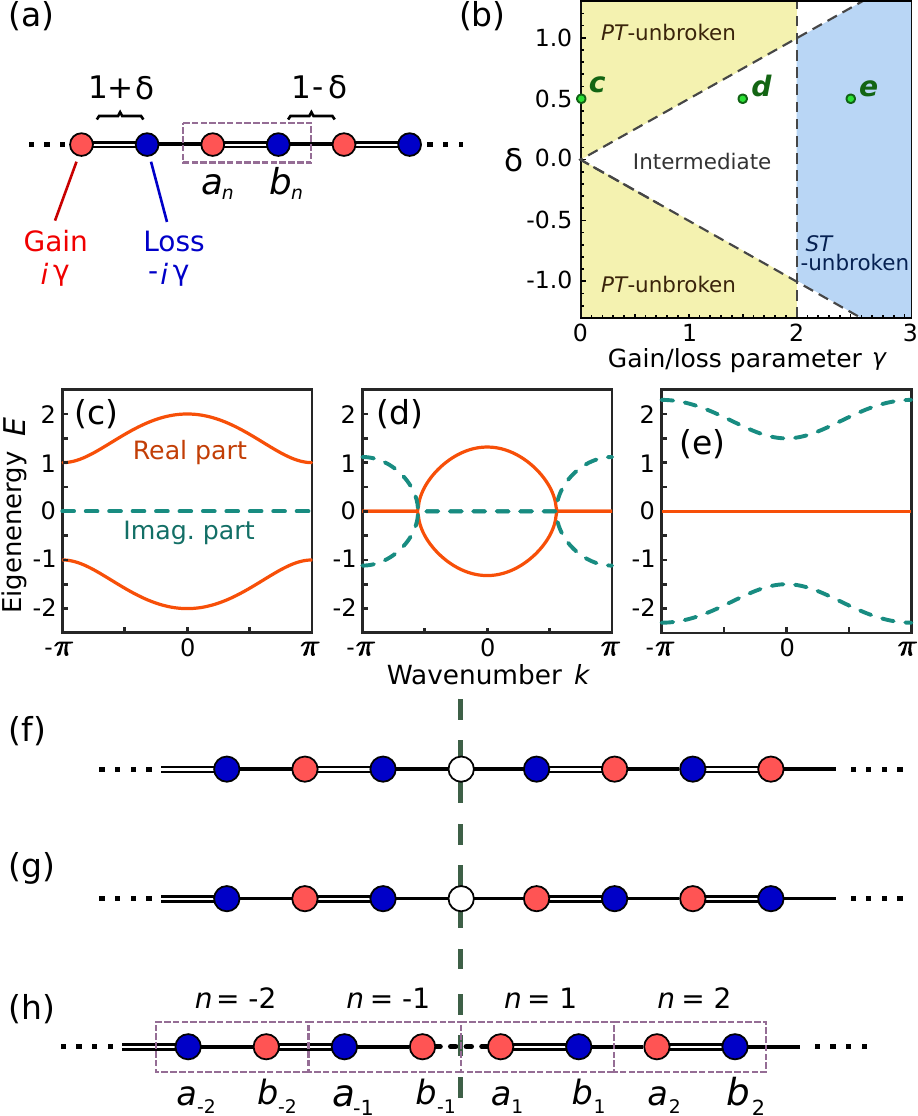}
  \caption{(a) The bulk cSSH chain.  Sites with gain ($i\gamma$) and loss ($-i\gamma$) are respectively indicated by red and blue circles, and couplings $1+ \delta$ and $1-\delta$ are indicated by double and single horizontal lines.  (b) Phase diagram of the bulk cSSH chain, which contains a $\mathcal{PT}$-unbroken phase where all Bloch state energies are real (yellow), an intermediate phase with both real and imaginary energies (white), and an $\mathcal{ST}$-unbroken phase with purely imaginary energies (blue).  (c)--(e) Complex bulk band structures for the cSSH chain, for the parameters indicated in (b) by the points labelled $c$, $d$, and $e$ respectively: $\delta = 0.5$ and (c) $\gamma =0$, (d) $\gamma = 1.5$, and (e) $\gamma = 2.5$.  Solid (dashed) curves show the real (imaginary) part of the eigenenergy $E$.  (f)--(h) cSSH chains with different lattice defects (vertical dashes).  (f) $\mathcal{P}$-preserving defect.  (g) $\mathcal{PT}$-preserving and $\mathcal{P}$-breaking defect.  (h) A defect that breaks both $\mathcal{P}$ and $\mathcal{PT}$, while reversing $\gamma$ and $\delta$ across the interface. }
  \label{fig1}
\end{figure}

The bulk cSSH lattice, depicted in Fig.~\ref{fig1}(a), consists of a chain of dimers with alternating coupling strengths $t \pm \delta$ between adjacent $a$ and $b$ sites, and alternating on-site gain/loss represented by imaginary on-site potentials $\pm i \gamma$.  The gain/loss averages to zero over the lattice.  The bulk Hamiltonian is
\begin{align}
  {\cal H}_{\mathrm{bulk}}
  &= \sum_n \Big[(t+\delta)\ket{a_n}\bra{b_n}+(t-\delta)\ket{a_n}\bra{b_{n-1}}
    +\text{h.c.}\Big] \nonumber\\
  &\quad + \sum_n\Big[i\gamma\ket{a_n}\bra{a_n}-i\gamma\ket{b_n}\bra{b_n}\Big],
  \label{Ham_real}
\end{align}
where $\ket{a_n}$ and $\ket{b_n}$ denotes the state on site $a$ and $b$, respectively, in the $n$-th unit cell.  The parameters $t$, $\delta$, and $\gamma$ are all real; we set $t=1$ as the energy unit.  When $\gamma = 0$, ${\cal H}_{\mathrm{bulk}}$ reduces to the SSH Hamiltonian~\cite{130704-1}.    Performing a Fourier decomposition yields the reduced Hamiltonian
\begin{equation}
  H_k= \begin{pmatrix} i\gamma & W_k  \\ W_{-k} & -i\gamma \end{pmatrix},
  \label{Ham_k}
\end{equation}
where $W_k=(1+\delta)+(1-\delta)e^{-ik}$ and $k$ is the crystal momentum, with the unit of length chosen so that the lattice constant is unity.  The resulting eigenvalue spectrum,
\begin{equation}
  E_{k,\pm}=\pm\sqrt{2(1+\delta^2)+2(1-\delta^2)\cos k-\gamma^2},
  \label{disp_rel_1}
\end{equation}
is shown in Fig.~\ref{fig1}(c)--(e) for different values of $\gamma$.

In the bulk, the SSH lattice has parity ($\mathcal{P}$), time-reversal ($\mathcal{T}$), and sublattice ($\mathcal{S}$) symmetries.  These symmetries are described in detail in Appendix~\ref{a-proof}.  For $\gamma \ne 0$, the cSSH model breaks $\mathcal{P}$, $\mathcal{S}$, and $\mathcal{T}$ individually, but retains two composite symmetries.  First, it is $\mathcal{ST}$ symmetric \cite{171007-1,180402-1,171207-7}, which implies that if $E_k$ is an eigenvalue, $-E_{-k}^*$ is also an eigenvalue \cite{180322-1,180306-1}; as discussed in Appendix~\ref{a-proof}, this symmetry is responsible for the flatness of the real part of the spectrum in the $\mathcal{ST}$-unbroken regime \cite{180322-1,180306-1, 180720-1}.  Second, it is $\mathcal{PT}$ symmetric, which implies that if $E_k$ is an eigenvalue, $E^*_k$ is also an eigenvalue \cite{101224-1, 180501-3, 180501-1}.  The phase diagram for the cSSH chain's bulk bandstructure is shown in Fig.~\ref{fig1}(b).  It is divided into three parts: (i) a $\mathcal{PT}$-unbroken phase where the Bloch states have real energies for all $k$, (ii) an intermediate phase where the energies are real for some ranges of $k$ and imaginary elsewhere, and (iii) an $\mathcal{ST}$-unbroken phase where the energies are purely imaginary for all $k$.

It is well known that when the SSH model ($\gamma = 0$) is gapped ($\delta \ne 0$), connecting two domains with opposite signs of $\delta$ leads to the emergence of a topological mid-gap defect state localized at the domain wall.  The energy of the defect state is pinned to exactly zero by the $\mathcal{S}$ symmetry~\cite{130704-1}.  In a similar vein, we can consider putting domain walls or defects in the cSSH model ($\gamma \ne 0$).  However, due to the presence of both alternating coupling strengths and alternating gain/loss in the cSSH model, there is some leeway in how the defect is defined.  Schomerus' original study of the cSSH model \cite{171007-1} used the configuration shown in Fig.~\ref{fig1}(f), with gain applied to the defect site.  In this case, the defect site can also be regarded as a domain wall, and the lattice is symmetric under a $\mathcal{P}$ operation across the defect site, whereas $\mathcal{PT}$ is broken \cite{171007-1,171212-2,180402-1}.  (Henceforth, we will let the bulk lattices on the two sides of the defect have the same values of $\delta$ and $\gamma$.)  This configuration has the notable feature that the bulk lattices on the two sides are ``incompatible'': they have different patterns of alternating $\delta$ and $\gamma$, and are related to each other by a swap of either $\delta$ or $\gamma$.  Subsequent studies \cite{171207-7, 180409-2, sshlaser1, 180419-3} have also considered the configuration shown in Fig.~\ref{fig1}(g); without gain or loss at the defect site, the overall lattice preserves $\mathcal{PT}$, although $\mathcal{P}$ is broken.  The lattices on either side of the defect are ``compatible'', in the sense that this configuration could be generated by inserting an additional site and link into a uniform cSSH lattice (similar to the original SSH case).

We will consider a third defect configuration, shown in Fig.~\ref{fig1}(h).  This has also been employed in a recent study by Yuce \cite{180614-1}.  Unlike the previous two cases, the domain wall can be regarded as lying between two lattice sites; this defect configuration can be generated by inserting an additional gain site and link with coupling strength 1 into a uniform cSSH lattice, similar to the defect of Fig.~\ref{fig1}(g) or the SSH model.  The motivation for studying this configuration is that in the $\delta \rightarrow 0$ limit, the lattice takes the form of a gain/loss dimer lattice \cite{180322-1} with a missing-site defect, which can also be regarded as a domain wall lying across a link.  Such a lattice supports ``non-Hermiticity-induced'' defect states, the implications of which will be discussed later.  As indicated in Fig.~\ref{fig1}(h), we label the unit cells by $n = 1, 2, \dots$ to the right of the defect, and $n = -1, -2, \dots$ to the left.  The cSSH sublattices on the two sides are related by a simultaneous swap of $\delta$ and $\gamma$.  The defect breaks both $\mathcal{P}$ and $\mathcal{PT}$, but the $\mathcal{ST}$ symmetry of the underlying cSSH lattice is preserved.  Consequently, eigenstates of the lattice must be either $\mathcal{ST}$ symmetric, or form pairs with eigenenergies $(E_D,-E_D^*)$.  For details, refer to Appendix \ref{a-proof}.  In the following, we focus on the case of $\gamma\ge 0$; the $\gamma\le 0$ case is just the time-reversed counterpart, with complex conjugated eigenenergies.

\section{Defect states}

\begin{figure}
  \includegraphics[width=0.99\linewidth]{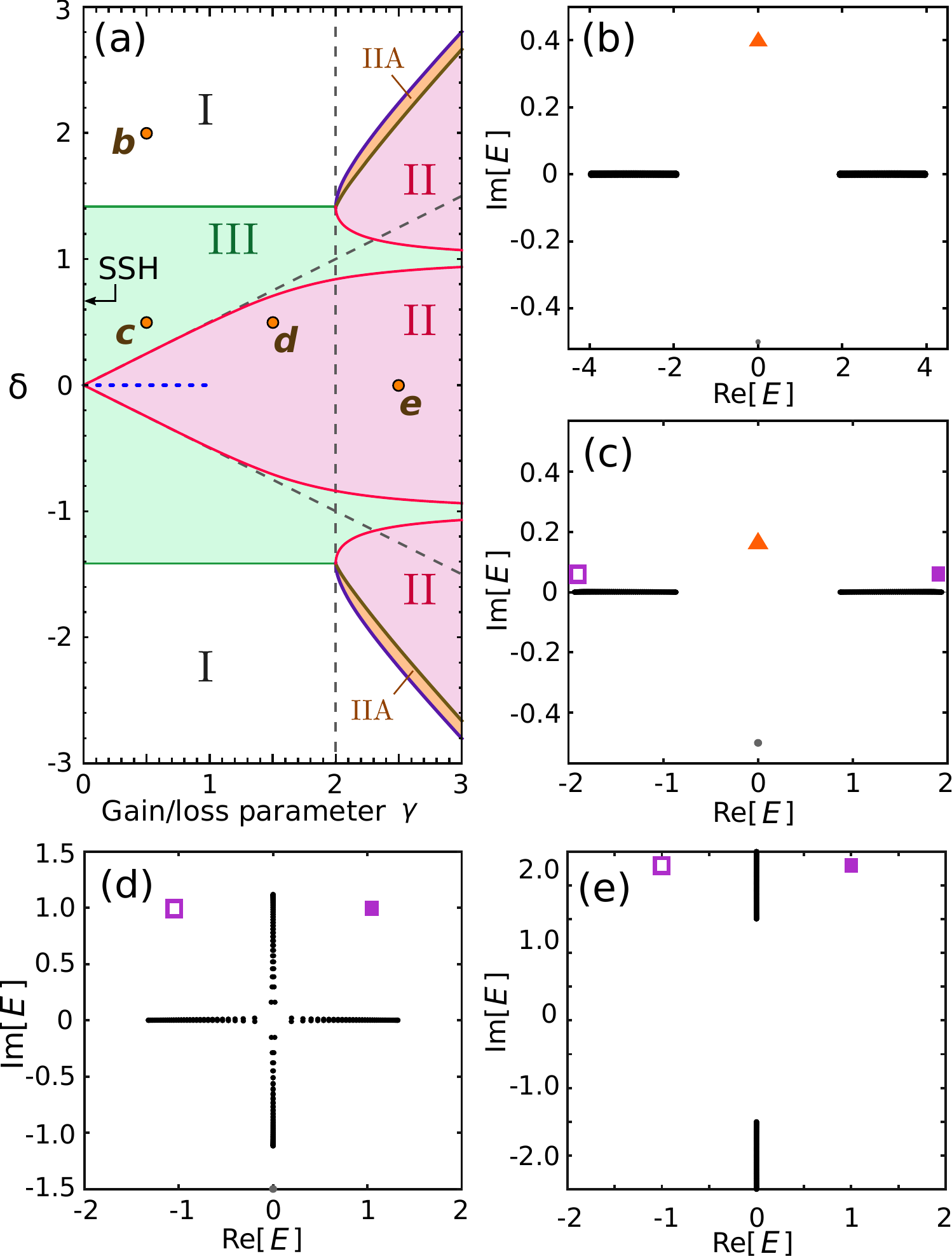}
  \caption{(a) Phase diagram for defect states of the cSSH lattice shown in Fig.~\ref{fig1}(h).  In the white regions (I), the lattice has one defect state, in the green regions (III) it has three defect states, and in the pink (II) and orange regions (IIA) it has two defect states.  The defect states in II are $\mathcal{ST}$-broken, and the defect states in IIA are $\mathcal{ST}$-symmetric.  The points labelled $b$, $c$, $d$, and $e$ indicate the parameters for the spectra plotted in (b)--(e).  The gray dashes are the phase boundaries of the bulk lattice bandstructure, corresponding to Fig.~\ref{fig1}(b).  The blue dots indicate the critical line segment ($\delta = 0$ and $0\le\gamma\le1$) over which there are no localized defect states.  (b)-(e) Complex eigenenergy spectra, calculated numerically for a finite lattice with 75 unit cells on each side of the defect, for (b) $\gamma = 0.5$ and $\delta = 2$, (c) $\gamma = 0.5$ and $\delta = 0.5$, (d) $\gamma = 1.5$ and $\delta = 0.5$, and (e) $\gamma = 2.5$ and $\delta = 0$.  Defect states continuable to SSH mid-gap states are shown as triangles, defect states continuable to non-mid-gap states as squares, bulk states as black circles, and edge states (due to the finite lattice size) as grey circles.}
  \label{fig2}
\end{figure}

We look for states that are exponentially localized to the defect, having the form
\begin{equation}
  \ket{\psi_D} = \sum_{\pm} \sum_{n > 0} \lambda^{n}
  \Big(\alpha_\pm |a_{\pm n}\rangle + \beta_\pm |b_{\pm n}\rangle\Big),
  \label{trial}
\end{equation}
with undetermined complex constants $\lambda$, $\alpha_\pm$, and $\beta_\pm$, constrained by $|\lambda| < 1$.  The lattice is assumed to be infinite, so the sum over $n$ extends to infinity.  The energy of the defect state is related to $\lambda$ by
\begin{equation}
  E_D^2+\gamma^2=2(1+\delta^2)+(1-\delta^2)(\lambda+\lambda^{-1}).
  \label{disp_2}
\end{equation}
This ansatz is applied to the lattice configuration shown in Fig.~\ref{fig1}(h), with coupling strength 1 on the defect link.  The solution is detailed in Appendix~\ref{a-derive}.  The resulting phase diagram for the defect states is shown in Fig.~\ref{fig2}(a).  (As previously discussed, we consider only $\gamma\ge0$.)

The left edge of this phase diagram ($\gamma = 0$) corresponds to an SSH model with a domain wall through a link, and topologically distinct configurations on either side.  This supports \textit{three} distinct defect state solutions.  One of them is the mid-gap defect state, which exists for all $\delta \ne 0$ and is pinned to energy $E_D^0 = 0$ by the $\mathcal{S}$ symmetry \cite{130704-1,111222-1} and topologically protected by a $\pi$ difference in the Zak phases calculated for two bulk lattices \cite{130107-1}.  The other two defect states, which exist for $0 < |\delta| < \sqrt{2}$, have eigenenergies
\begin{equation}
  E_D^{\pm} = \pm \sqrt{4+\delta^4},
\end{equation}
which respectively lie above and below the bulk energy bands.  These defect states do not lie in a band gap, and are usually regarded as being topologically trivial in the Hermitian sense, because the Zak phases related to the non-mid-gap states have no difference for the two configurations on both sides of the domain wall.  In the limit $\delta = \pm 1$, the three defect states reduce to the eigenstates of a trimer with coupling strengths of 1 and 2 on its two links, whose eigenenergies are 0 and $\pm\sqrt{5}$.

Suppose $0 < |\delta|<\sqrt{2}$, for which the SSH chain has three defect states.  As we gradually increase $\gamma$ from zero, all three defect states evolve continuously into exponentially-localized defect states of the non-Hermitian lattice, and their energies $\{E_D^0, E_D^\pm\}$ become complex.  The topological mid-gap state becomes an unpaired $\mathcal{ST}$-symmetric defect state with imaginary $E_D^0$.  The two non-mid-gap states become a pair of defect states that individually break the $\mathcal{ST}$ symmetry, and map to each other under $\mathcal{ST}$, satisfying $E_D^+ = - (E_D^-)^*$.  Within the regions labelled III in Fig.~\ref{fig2}(a), the lattice supports three distinct and well-defined defect states.  Note that the boundary of these regions lie close to, but outside, the phase boundaries of the bulk bandstructure's $\mathcal{PT}$-symmetric phase [Fig.~\ref{fig1}(b)].  Fig.~\ref{fig2}(c) shows a typical complex energy spectrum in domain III, calculated for a large but finite lattice.  (We emphasize, however, that the phase boundaries in Fig.~\ref{fig2}(a) were derived for infinite lattices, with and without the domain wall.)  Note that the two $\mathcal{ST}$-broken defect states (blue squares) have values of $\mathrm{Re}(E_D^\pm)$ overlapping with the real bulk energy bands, but are nonetheless exponentially localized to the defect.  This is reminiscent of the phenomenon of ``bound states in the continuum'' in Hermitian systems \cite{180703-2,180703-1}, whose realization in non-Hermitian systems has recently been discussed by several authors \cite{180703-4,180703-3,180703-5}.

As $\gamma$ is further increased, the unpaired $\mathcal{ST}$-symmetric defect state abruptly disappears.  In the domain labelled II in Fig.~\ref{fig2}(a), the system contains only two defect states (the $\mathcal{ST}$-broken pair).  The disappearance of the $\mathcal{ST}$-symmetric defect state occurs via a divergence in its exponential decay constant (i.e., $|\lambda| \ge 1$ in the solution to the Schr\"odinger equation with the ansatz \eqref{trial}), which coincides with the merging of $E_D^0$ into the continuum of bulk state eigenenergies.  Since $E_D^0$ is imaginary, this can only happen outside the $\mathcal{PT}$-symmetric phase of the bulk bandstructure, where the bulk spectrum is at least partially imaginary---hence the relationship between the defect state phase boundary and bulk phase boundary in Fig.~\ref{fig2}(a).  Further details about the disappearance of the $\mathcal{ST}$-symmetric defect state are given in Appendix~\ref{a-derive}.

Fig.~\ref{fig2}(d) and (e) shows the complex energy spectra at two points in domain II.  In Fig.~\ref{fig2}(d), the bulk is in the intermediate phase, and its energies lie partly on the real line and partly on the imaginary line; the defect state eigenenergies $E_D^\pm$ stand apart from the bulk energies in the complex plane, but their real parts can be embedded in the $\mathrm{Re}(E)$ continuum.  In Fig.~\ref{fig2}(e), the bulk is in the $\mathcal{ST}$-symmetric phase and all of its energies are imaginary, whereas $\mathrm{Re}(E_D^\pm) \ne 0$.

The case of $\delta = 0$ deserves special attention.  For $\gamma = 0$, this is just an undimerized chain, with no defect states.  More interestingly, there are no defect states over the finite range $0 \le \gamma \le 1$.  Only for $\gamma > 1$ do the pair of $\mathcal{ST}$-broken defect states appear, described by
\begin{align}
  \alpha_- &= \beta_+ = \pm \lambda,\;\;\;
  \alpha_+ = \beta_- = 1, \\
  \lambda &= \pm \left(\sqrt{1-\gamma^2}-i\gamma\right). \label{d0lambda}
\end{align}
The corresponding eigenenergies are $E_D^\pm =\sqrt{1-\gamma^2}\pm 1$.  Within the range $0 \le \gamma \le 1$, Eq.~\eqref{d0lambda} states that $|\lambda| = 1$, so the states are not localized.  This results in a critical line segment in the phase diagram, indicated by the blue dots in Fig.~\ref{fig2}(a), on which no localized defect states exist.  The end of the line segment (at $\delta = 0,\,\gamma = 1$) is an exceptional point for Eq.~\eqref{d0lambda}.

Malzard, Poli, and Schomerus \cite{171009-1} have recently drawn attention to a class of intrinsically non-Hermitian defect states that (i) are not present in the Hermitian limit, and (ii) appear when a non-Hermiticity parameter exceeds a certain nonzero magnitude.  They argued that such defect states may be considered ``topologically protected'', in the sense that they are associated with non-Hermitian spectral phases bounded by exceptional points related to $\mathcal{PT}$ symmetry breaking. The defect states of our $\delta = 0$ lattice behave similarly, and in fact we show in Appendix \ref{a-eqv} that the $\delta = 0$ lattice is a particular limit of the model in Ref.~\onlinecite{171009-1}.  However, the present analysis reveals qualifications to regarding these as topological defect states.  In the $0 \le \gamma \le 1$ range, the defect states are indeed absent for $\delta = 0$, but instantly re-appear when an infinitesimal $\delta$ is introduced (which causes $|\lambda|$ to drop below 1).  Moreover, the defect states themselves are continuable to the non-mid-gap defect states of the SSH lattice, which are not topologically protected in the Hermitian sense.

\section{$\mathcal{ST}$-breaking of defect states}

\begin{figure}
  \includegraphics[width=0.9\linewidth]{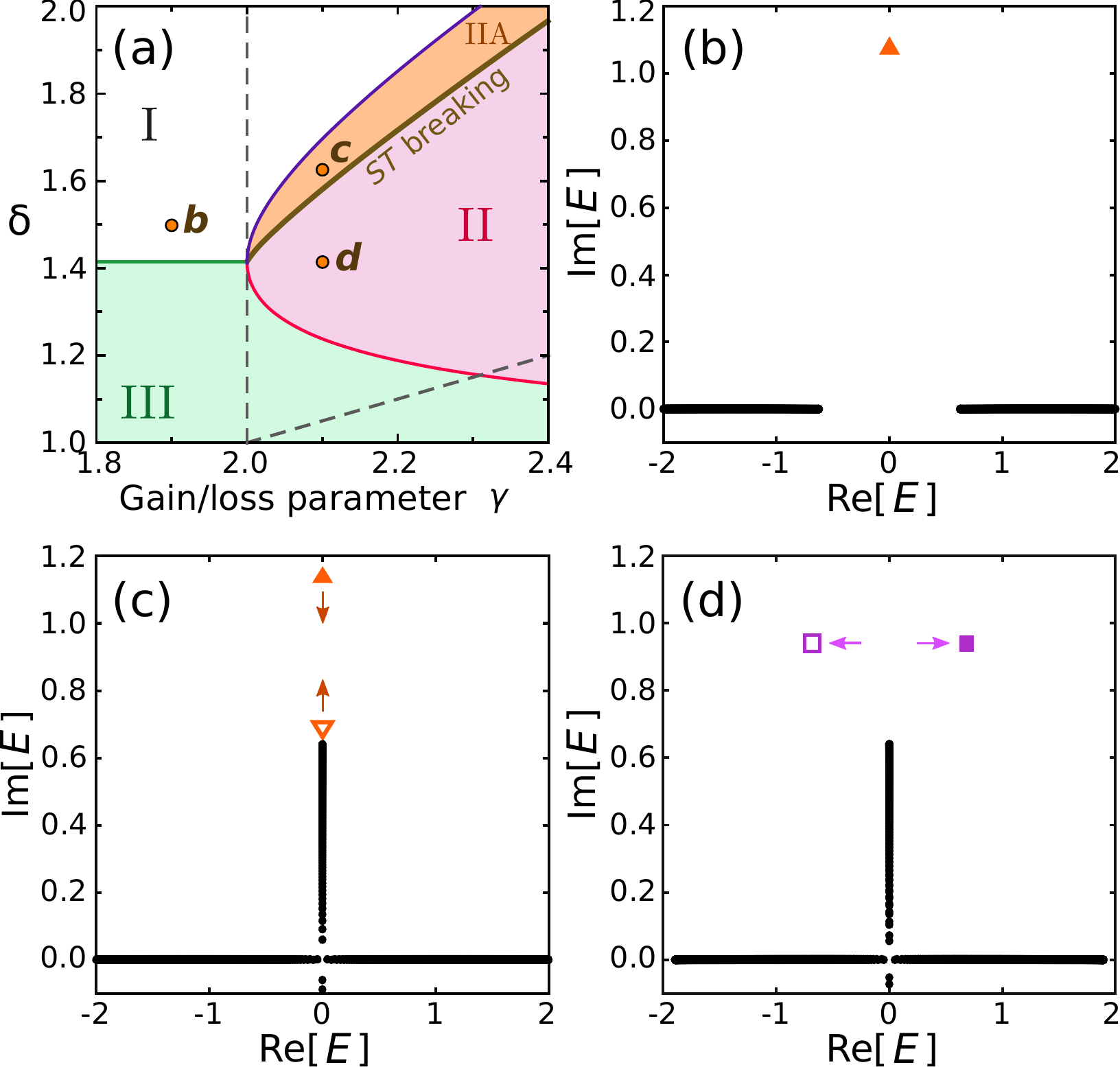}
  \caption{(a) Close-up view of the defect state phase diagram near the $\mathcal{ST}$-breaking line.  The region labels have the same meanings as in Fig.~\ref{fig2}(a).  The points labelled $b$, $c$, and $d$ indicate the parameters for the spectra plotted in (b)--(d). (b)-(d) Complex eigenenergy spectra, calculated numerically for a finite lattice with 500 unit cells on each side of the defect, for (b) $\gamma = 1.9$ and $\delta = 1.5$, (c) $\gamma = 2.1$ and $\delta = 1.625$, and (d) $\gamma = 2.1$ and $\delta = \sqrt{2}$. The $\mathcal{ST}$-symmetric defect states states are indicated by triangles; the $\mathcal{ST}$-broken defect states are indicated by squares.  The arrows in (c) and (d) indicate the direction of motion of the defect state eigenvalues as $\delta$ decreases from point $c$ to point $d$ in (a).}
  \label{fig3}
\end{figure}


Fig.~\ref{fig3}(a) shows a close-up view of the phase diagram for $\delta \sim \sqrt{2}$.  For $|\delta|>\sqrt{2}$ and $\gamma = 0$, the SSH lattice has a single defect state (the mid-gap defect state).  As we increase $\gamma$ from zero, keeping $\delta$ fixed, the eigenvalue $E_D^0$ moves up the imaginary axis, as shown in Fig.~\ref{fig3}(b).

As the system enters the region labelled IIA, another $\mathcal{ST}$-symmetric defect state emerges from the continuum, as shown in the complex spectrum plotted in Fig.~\ref{fig3}(c). Hence, in this region there are \textit{two} $\mathcal{ST}$-symmetric states that are localized to the defect, a phenomenon with no counterpart in the Hermitian SSH model.

As the system moves from region IIA to region II, the two imaginary energies approach each other, meet, then move off the imaginary axis to either side, as shown in Fig.~\ref{fig3}(d).  The transition line, shown as a thick brown line in Fig.~\ref{fig3}(a), signifies a $\mathcal{ST}$-breaking transition.  In region II, as we have previously discussed, the system has two $\mathcal{ST}$-broken defect states, which are continuable to topologically trivial defect states of the SSH chain.

Thus, there appear to be two distinct ways for the SSH mid-gap defect state to disappear from the non-Hermitian lattice.  The first is to merge into the continuum; the second is to undergo a $\mathcal{ST}$-breaking transition with another $\mathcal{ST}$-symmetric defect state emerging from the continuum.

\begin{figure}
  \includegraphics[width=0.99\linewidth]{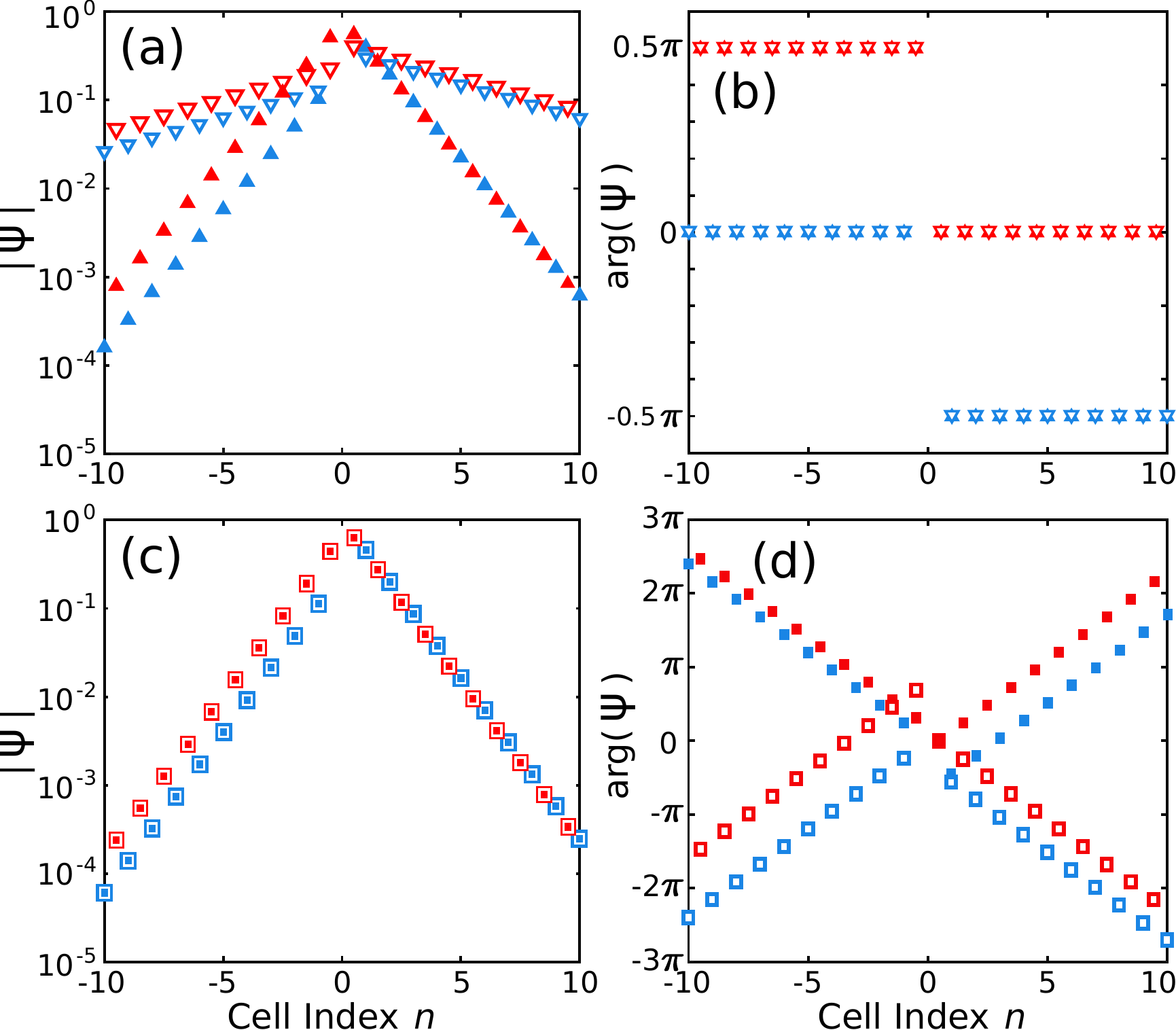}
  \caption{Magnitudes and phases of the cSSH defect state wavefunctions, before and after an $\mathcal{ST}$-breaking transition.  In (a) and (b), the lattice parameters are $\gamma = 2.1$ and $\delta = 1.625$, the same as in Fig.~\ref{fig3}(c); both eigenstates are $\mathcal{ST}$-symmetric.  The up-pointing triangles show the state that evolved from the SSH mid-gap defect state; the down-pointing triangles show the additional $\mathcal{ST}$-symmetric defect state which emerged from the continuum.  In (c) and (d), the lattice parameters are $\gamma = 2.1$ and $\delta = \sqrt{2}$, the same as in Fig.~\ref{fig3}(d); both eigenstates are $\mathcal{ST}$-broken, and are related by $\mathcal{ST}$ operation.  The filled squares show the state with $\mathrm{Re}(E) > 0$ and the hollow squares show the state with $\mathrm{Re}(E) < 0$.  Red (blue) symbols indicate gain (loss) sites.  The gauge is fixed by setting the phase of the first site to the right of the domain wall to zero.}
  \label{fig4}
\end{figure}

Fig.~\ref{fig4} shows the eigenstate magnitudes and phases of the defect states on either side of the $\mathcal{ST}$-breaking transition.  On one side of the transition, the two $\mathcal{ST}$-symmetric states have different intensity profiles with different localization lengths; moreover, the phases in the gain and loss sites differ by $\pi/2$, which is a characteristic feature of unbroken $\mathcal{ST}$ symmetry (see Appendix~\ref{a-derive}).  On the other side of the transition, the two $\mathcal{ST}$-broken eigenstates have identical intensity profiles, and the phases on the right (left) side of the domain wall are symmetric with respect to $0$ ($\pi/2$) for the gain sites and $-\pi/2$ ($0$) for the loss sites; these features arise from the fact that the two states are related by the $\mathcal{ST}$ operation.

\section{Discussion}
\label{conclude}

The cSSH model may be regarded as the simplest one-dimensional non-Hermitian model with a clear link to Hermitian concepts of band topology.  In this paper, we have examined the conditions under which a cSSH lattice supports exponentially localized defect states.  Previous papers on the subject have focused on the simplest case of a single non-Hermitian defect state that is $\mathcal{ST}$-symmetric, whose energy has exactly zero real part.  Such a defect state has a clear connection to the physics of topological states: it is continuable, in the Hermitian limit, to the SSH model's well-known topological defect state \cite{171007-1, 171212-2,171207-7,180402-1, 180409-1,180409-2,180419-3, 180719-2}.

Our study has revealed a richer variety of behaviors.  In particular, the cSSH model has defect states that are $\mathcal{ST}$-broken, with energies having non-zero real parts.  Although these are continuable to the SSH model's ``trivial'' defect states, they play an interesting role in the cSSH model.  In some parameter regimes, a pair of $\mathcal{ST}$-broken defect states can co-exist with an $\mathcal{ST}$-symmetric defect state.  Alternatively, an $\mathcal{ST}$-symmetric defect state can coalesce with another $\mathcal{ST}$-symmetric defect state emerging out of the continuum, turning into an $\mathcal{ST}$-broken pair.  This is an inherently non-Hermitian phenomenon that lacks any analogue in the SSH model.

We have focused on the specific defect configuration of Fig.~\ref{fig1}(h), with coupling strength 1 on the defect link.  If the couping strength is not 1, the phase diagram for the defect states is qualitatively similar, though the positions of the phase boundaries are shifted, and the critical line segment at $\delta = 0$ is not present.  For the alternative configurations shown in Figs.~\ref{fig1}(f)-(g), there exist similar combinations of $\mathcal{ST}$-broken and $\mathcal{ST}$-unbroken defect states, but the phase diagrams are different.  For the configuration of Fig.~\ref{fig1}(f), a single $\mathcal{ST}$-unbroken defect state exists for $\delta > 0$, whereas for $\delta > 0$ there are three defect states (one $\mathcal{ST}$-unbroken and two $\mathcal{ST}$-broken, or three $\mathcal{ST}$-unbroken).  For the configuration of Fig.~\ref{fig1}(g), there is a phase with no defect states, similar to the critical line in Fig.~\ref{fig2}(a); since this configuration is also $\mathcal{PT}$ symmetric, its $\mathcal{ST}$-broken defect state pairs have real energies.

\begin{acknowledgments}
  We are grateful to H.~Schomerus and D.~Leykam for helpful discussions.  This work was supported by the Singapore MOE Academic Research Fund Tier 2 Grant MOE2015-T2-2-008 and the Singapore MOE Academic Research Fund Tier 3 Grant MOE2016-T3-1-006.
\end{acknowledgments}

\appendix
\section{Symmetries of cSSH model}
\label{a-proof}

Consider a one-dimensional discrete lattice with unit cells labelled by an integer $n$, and $N_s$ sites in each unit cell.  Let $\psi_n$ be a column vector consisting of the $N_s$ annihilation operators in unit cell $n$.  In this general context, we can define the time-reversal operator ($\mathcal{T}$), charge-conjugation operator (${\cal C}$, also called the particle-hole operator), sublattice operator (${\cal S}$, also called the chiral operator), and parity operator (${\cal P}$) as follows \cite{120316-1}:
\begin{align}
  \begin{aligned}
    {\cal T}\psi_{n}{\cal T}^{-1} &= U_{\cal T}\psi_{n},
    \,~~{\cal C}\psi_{n}{\cal C}^{-1}=U_{\cal C}^{*}\psi_{n}^{\dag}, \\
    {\cal S}\psi_{n}{\cal S}^{-1} &= U_{\cal S}\psi_{n},~~{\cal P}\psi_{n}{\cal P}^{-1}=U_{\cal P}\psi_{-n}.
  \end{aligned}
\end{align}
Here, $U_{\cal T,C,S,P}$ are unitary matrices, and the parity operation is taken around the origin.

A system is said to be $\mathcal{T}$, $\mathcal{P}$, and $\mathcal{S}$ symmetric if its lattice Hamiltonian $\mathcal{H}$ satisfies, respectively,
\begin{equation}
  [\mathcal{T},\mathcal{H}] = 0, \;\;
  [\mathcal{P},\mathcal{H}] = 0, \;\;
  \{\mathcal{S},\mathcal{H}\} = 0.
\end{equation}
These definitions apply to translationally invariant lattices as well as lattices with defects.

The bulk SSH model satisfies all three symmetries, with the matrix representations
\begin{equation}
  U_\mathcal{T}=I, \;\; U_\mathcal{P}=\sigma_x, \;\;
  U_\mathcal{S}=\sigma_z. \label{representations}
\end{equation}
Using these same matrix representations, the bulk cSSH model breaks $\mathcal{T}$, $\mathcal{S}$, and $\mathcal{P}$ individually, but preserves $\mathcal{PT}$ and $\mathcal{ST}$.

For any infinite translationally invariant one-dimensional lattice, the Hamiltonian has the form
\begin{equation}
  {\cal H} =\sum_k(\psi_k^\dag)^T H_k \psi_k,
\end{equation}
where $\psi_{k}=N^{-1/2} \sum_{k} \exp(-iknd) \psi_{n}$.  For simplicity, we set the lattice constant $d$ to unity.  In terms of these Bloch-state operators, the above symmetry operators have the following form:
\begin{align}
  \begin{aligned}
    {\cal T}\psi_{k}{\cal T}^{-1} &= U_{\cal T}\psi_{-k},
    ~~{\cal C}\psi_{k}{\cal C}^{-1}=U_{\cal C}^{*}\psi_{-k}^{\dag}, \\
    {\cal S}\psi_{k}{\cal S}^{-1} &= U_{\cal S}\psi_{k},
    \;~~{\cal P}\psi_{k}{\cal P}^{-1}=U_{\cal P}\psi_{-k}.
  \end{aligned}
\end{align}
Thus,
\begin{align}
  {\cal T}{\cal H}{\cal T}^{-1} &=\sum_k(\psi_k^\dag)^T (U_{{\cal T}}^\dag H^*_{-k} U_{{\cal T}})\psi_k, \label{THT} \\
  {\cal C}{\cal H}{\cal C}^{-1} &= \text{Tr}({\cal H})+\sum_k(\psi_k^\dag)^T (\pm U_{{\cal C}}^\dag H^T_{-k} U_{{\cal C}})\psi_k, \label{CHC}\\
  {\cal S}{\cal H}{\cal S}^{-1} &= \sum_k(\psi_k^\dag)^T (U_{{\cal S}}^\dag H_{k} U_{{\cal S}})\psi_k, \\
  {\cal P}{\cal H}{\cal P}^{-1} &= \sum_k(\psi_k^\dag)^T (U_{{\cal P}}^\dag H_{-k} U_{{\cal P}})\psi_k. \label{PHP}
\end{align}
In Eq.~\eqref{CHC}, $\pm$ hold for bosons and fermions respectively.  Notably, the relation ${\cal CHC}^{-1}=\text{Tr}(\mathcal{H})\pm {\cal H}$ is always satisfied with $U_{\cal C}=I$ (identity matrix); hence, $H_{-k}^T=H_k$, which implies the general eigenenergy pair $(E_k,E_{-k})$ at each crystal momentum $k$, i.e., the band structure is mirror-symmetric with respect to $k=0$.

For the bulk SSH lattice, Eqs.~\eqref{THT}--\eqref{PHP} lead respectively to the relations
\begin{align}
  H_{-k}^* &= H_k \\
  \sigma_x H_{-k}\sigma_x &= H_{k} \\
  \sigma_z H_{k}\sigma_z &= -H_k.
\end{align}
Hence, the eigenenergies appear in pairs, $(E_k,E_{-k})$ and $(E_k,-E_{k})$, for each $k$.  The band diagram is mirror-symmetric around both $k=0$ and $E=0$.

In the bulk cSSH lattice, the $\mathcal{PT}$ and $\mathcal{ST}$ symmetries respectively imply that
\begin{align}
  \sigma_x H^*_k \sigma_x &= H_k \\
  \sigma_z H_{-k}^*\sigma_z &= -H_k.
\end{align}
As a consequence, if there is a bulk state of energy $E_k$, there must exist a bulk state of energy $E_k^*$ (due to $\mathcal{PT}$), and a bulk state of energy $-E_{-k}^*$ (due to $\mathcal{ST}$).  The former ensures that in the $\mathcal{PT}$-unbroken phase, the bands are purely real.  The latter can be combined with the definition of $\mathcal{C}$ to yield $\{{\cal H,STC}\}=\mp1$, and hence
\begin{eqnarray}
  \sigma_z H_{k}^\dag \sigma_z=-H_k.
\end{eqnarray}
This ensures that if there is a bulk state of energy $E_k$, there must exist a bulk state of energy $-E_k^*$.  In the ${\cal ST}$-unbroken phase, the bands are purely imaginary \cite{180402-1}.

	
For the cSSH chains with domain walls discussed in the main text, the overall ${\cal PT}$ symmetry is broken but ${\cal ST}$ is preserved [using the representations \eqref{representations}].  Hence, defect states must either be $\mathcal{ST}$-unbroken (and hence have purely imaginary eigenenergies), or appear in pairs with eigenenegies $\{E_D,-E_D^*\}$ and eigenstates related to each other by $\mathcal{ST}$:
\begin{align}
  \begin{aligned}
    \mathcal{H(ST}\ket{\psi_D}) &= -\mathcal{(ST)H}\ket{\psi_D}) \\
    &= -E^*_D\mathcal{(ST}\ket{\psi_D}).
  \end{aligned}
\end{align}

\section{Solving for defect states}
\label{a-derive}

We find defect state solutions by substituting the ansatz \eqref{trial} into the time-independent Schr\"{o}dinger equation $(\mathcal{H}_+ + \mathcal{H}_-) \ket{\psi_D}=E_D\ket{\psi_D}$, where $H_+$ and $H_-$ are the Hamiltonians for semi-infinite lattices to the right ($n>0$) and left ($n<0$) of the domain wall.  This yields
\begin{align}
  H_\pm \psi_\pm &= E_D\psi_\pm, \label{Hpm}\\
  (1-\delta)\alpha_-+\alpha_+ &= (E_D-i\gamma)\beta_-, \label{eqset2}\\
  (1+\delta)\beta_++\beta_- &= (E_D-i\gamma)\alpha_+, \label{eqset3}
\end{align}
where 
\begin{align}
  H_\pm &= \begin{pmatrix}
    \pm i\gamma & W_{\pm\delta}(\lambda^{\mp1}) \\
    W_{\pm\delta}(\lambda^{\pm 1}) & \mp i\gamma \end{pmatrix} \\
  W_\delta(\lambda) &= (1+\delta)+(1-\delta)\lambda, \\
  \psi_{\pm} &= (\alpha_{\pm}, \;\beta_{\pm})^T.
\end{align}
From Eq.~\eqref{Hpm}, we derive Eq.~\eqref{disp_2}.  Without loss of generality, we can take
\begin{equation}
  \psi_-= \begin{pmatrix}
    E_D-i\gamma\\
    W_{-\delta}(\lambda^{-1})
  \end{pmatrix},\;\;\;
  \psi_+=\eta \begin{pmatrix}
    E_D+ i\gamma\\
    W_\delta(\lambda)
  \end{pmatrix}.
  \label{coeff}
\end{equation}
Eqs.~\eqref{eqset2} and \eqref{eqset3} describe the wave-matching conditions around the domain wall. Substituting Eq.~\eqref{coeff} into them, and using Eq.~\eqref{disp_2}, gives
\begin{align}
  \begin{aligned}
    E_D^2+\gamma^2 &= 2(1+\delta^2)+(1-\delta^2)(\lambda+\lambda^{-1}), \\
    \lambda &= \frac{(E_D-i\gamma)(1-\delta^2)}{E_D+i\gamma},
    \;\;\; \eta=\frac{1}{1-\delta}.
  \end{aligned}
  \label{eqs_defect}
\end{align}
We then search analytically or numerically for solutions satisfying $|\lambda| < 1$, corresponding to exponentially localized defect states.  From \eqref{eqs_defect}, we can see that if $(E_D,\lambda)$ is a solution, $(-E_D^*,\lambda^*)$ is also a solution.  Specifically, the $\mathcal{ST}$ couterpart satisfies
\begin{equation}
  U_\mathcal{S}\mathcal{K}(\psi_\pm\lambda^n)= \sigma_z(\psi_\pm\lambda^n)^*,
  \label{ST-coeff}
\end{equation}
where $\mathcal{K}$ is the complex conjugation operator.  This means that if the $\mathcal{ST}$-paired defect states are two distinct solutions, they have the same intensity profiles and the phases are symmetric with respect to $0\,(\pi/2)$ for the left (right) sites in each unit cell, which is precisely the behavior observed in Fig.~\ref{fig4}.

\begin{figure}
  \includegraphics[width=0.99\linewidth]{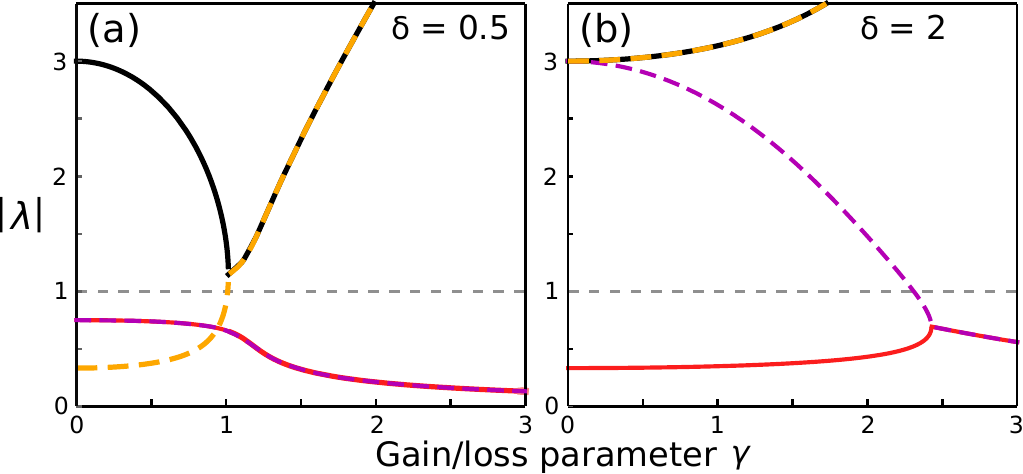}
  \caption{Variation of $|\lambda|$ with $\gamma$ for (a) $\delta=0.5$ and (b) $\delta=2$.}
  \label{evolution}
\end{figure}

The number of defect state solutions is determined by the number of solutions with $|\lambda| < 1$.  Eliminating $E_D$ in \eqref{eqs_defect} yields a fourth-order polynomial in $\lambda$.  In Fig.~\ref{evolution}, we plot $|\lambda|$ versus $\gamma$ for two different values of $\delta$.  If a root crosses the $|\lambda| = 1$ line, it corresponds to a defect state appearing out of, or disappearing into, the continuum.

Analytic solutions can be found for some special cases.  First, in the Hermitian limit ($\gamma=0$), there is a mid-gap defect state of the form
\begin{equation}
  \psi_- = \begin{pmatrix} -1\\ 0 \end{pmatrix}, \;\;
  \psi_+ = \begin{pmatrix} 1-\delta\\ 0 \end{pmatrix}, \;\;
  \lambda=-\frac{1+\delta}{1-\delta}
  \label{trimer_neg}
\end{equation}
for $\delta<0$, and
\begin{equation}
  \psi_- = \begin{pmatrix} 0\\ 1+\delta \end{pmatrix}, \;\;
  \psi_+ = \begin{pmatrix} 0\\ -1 \end{pmatrix}, \;\;
  \lambda=-\frac{1-\delta}{1+\delta}
  \label{trimer_pos}
\end{equation}
for $\delta>0$.  The energy is pinned to $E_D=0$ by the $\mathcal{S}$ symmetry \cite{130704-1,111222-1}, and the existence of this defect state is tied to the topologically distinct configurations on both sides of the domain wall as characterized by a $\pi$ difference in Zak phases \cite{130107-1}.
Additionally, when $0<|\delta|<\sqrt{2}$, there are two non-mid-gap defect state solutions:
\begin{align}
  \psi_- &= \begin{pmatrix}
    \pm(1-\delta)\sqrt{4+\delta^4}\\
    2-2\delta+\delta^2
  \end{pmatrix}, \;\;\;
  \psi_+ = \begin{pmatrix}
    \pm\sqrt{4+\delta^4}\\
    2-\delta^2+\delta^3
  \end{pmatrix} \nonumber \\
  \lambda &= 1-\delta^2,~~E_D=\pm \sqrt{4+\delta^4}.
\end{align}

Another analytic solution can be obtained when the inter-site couplings are uniform ($\delta=0$).  In this case, there can be a pair of defect state solutions of form
\begin{align}
  \begin{aligned}
    \psi_- &= \begin{pmatrix} \pm\lambda\\ 1 \end{pmatrix}, \;\;
    \psi_+ = \begin{pmatrix} 1\\\pm\lambda \end{pmatrix}, \\
    \lambda&=\pm(\sqrt{1-\gamma^2}-i\gamma),~E_D=\sqrt{1-\gamma^2}\pm 1.
  \end{aligned}
\end{align}
This is valid only for $\gamma > 1$; for $0 < \gamma < 1$, the defect state is not localized since $|\lambda| = 1$.  The two eigenenergies have the same imaginary part and opposite real parts, due to the $\mathcal{ST}$ antisymmetry.

Finally, for $\delta=\pm 1$, there is an isolated trimer at the defect, and we can determine the three eigenvalues
\begin{eqnarray}
  (E_D+i\gamma)(E_D-i\gamma)^2=5E_D-3i\gamma.
\end{eqnarray}
In the Hermitian case, the roots are $E_D= \{0,~\pm\sqrt{5}\}$.

\section{Relation to the model of Ref.~\cite{171009-1}}\label{a-eqv}

In this appendix, we show that the uniform-coupling ($\delta=0$) case of the cSSH lattice in the main text is related to the two-chain model in Ref.~\cite{171009-1}.

\begin{figure}
  \includegraphics[width=0.65\linewidth]{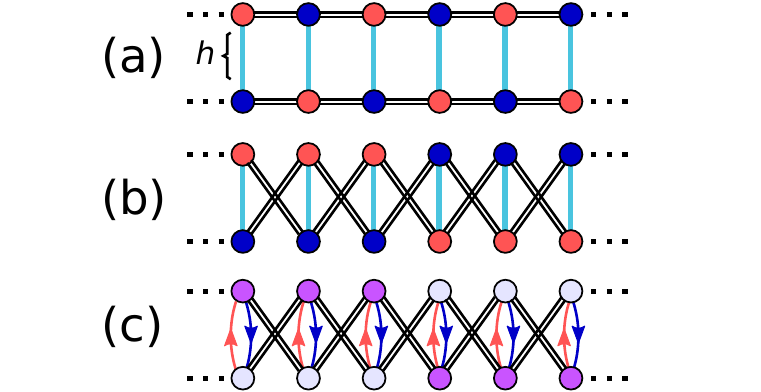}
  \caption{Mapping between (a) a pair of coupled uniform-coupling ($\delta = 0$) cSSH chains and (c) the two-chain model of Ref.~\onlinecite{171009-1}, where the purple (white) sites represent counterclockwise (clockwise) resonator modes, and the red and blue arrows represent asymmetric internal scattering processes.}
  \label{f-Schomerus}
\end{figure}

Let us couple a $\delta=0$ cSSH lattice to its time-reversed counterpart, transversely and site-by-site, to produce a two-chain lattice, as shown in Fig.~\ref{f-Schomerus}(a).  The inter-chain coupling is a new parameter denoted by $h$.  Next, we exchange the positions of even (odd) sites between two chains, as shown in Fig.~\ref{f-Schomerus}(b).  This causes the unit cell to shrink to a single column.  We then perform the following pseudo-rotation in each unit cell:
\begin{equation}
  \begin{pmatrix}
    \ket{u'_n}\\
    \ket{v'_n}
  \end{pmatrix}
  = {\cal U}
  \begin{pmatrix}
    \ket{u_n}\\
    \ket{v_n}
  \end{pmatrix},
\end{equation}
where ${\cal U}=e^{i\frac{\pi}{4}\sigma_x}=\frac{1}{\sqrt{2}}(1+i\sigma_x)$, and $\ket{u_n}$ and $\ket{v_n}$ are states localized to sites in the upper and lower chains in the $n$-th unit cell. In this new basis, the model is identical to that of Ref.~\cite{171009-1}, as shown in Fig.~\ref{f-Schomerus}(c), with the parameter correspondence
\begin{equation}
  A=h+\gamma,~~B=h-\gamma,~~W=t,	
\end{equation}
where $\{A,B,W\}$ are the parameters defined in Ref.~\onlinecite{171009-1}.

The uniform-coupling model discussed in the main text corresponds to $h=0$, which is the diagonal line in the phase diagram Fig. 2(c) in Ref.~\cite{171009-1} (i.e., $A=-B$).

%


\begin{thebibliography}{46}%
\makeatletter
\providecommand \@ifxundefined [1]{%
 \@ifx{#1\undefined}
}%
\providecommand \@ifnum [1]{%
 \ifnum #1\expandafter \@firstoftwo
 \else \expandafter \@secondoftwo
 \fi
}%
\providecommand \@ifx [1]{%
 \ifx #1\expandafter \@firstoftwo
 \else \expandafter \@secondoftwo
 \fi
}%
\providecommand \natexlab [1]{#1}%
\providecommand \enquote  [1]{``#1''}%
\providecommand \bibnamefont  [1]{#1}%
\providecommand \bibfnamefont [1]{#1}%
\providecommand \citenamefont [1]{#1}%
\providecommand \href@noop [0]{\@secondoftwo}%
\providecommand \href [0]{\begingroup \@sanitize@url \@href}%
\providecommand \@href[1]{\@@startlink{#1}\@@href}%
\providecommand \@@href[1]{\endgroup#1\@@endlink}%
\providecommand \@sanitize@url [0]{\catcode `\\12\catcode `\$12\catcode
  `\&12\catcode `\#12\catcode `\^12\catcode `\_12\catcode `\%12\relax}%
\providecommand \@@startlink[1]{}%
\providecommand \@@endlink[0]{}%
\providecommand \url  [0]{\begingroup\@sanitize@url \@url }%
\providecommand \@url [1]{\endgroup\@href {#1}{\urlprefix }}%
\providecommand \urlprefix  [0]{URL }%
\providecommand \Eprint [0]{\href }%
\providecommand \doibase [0]{http://dx.doi.org/}%
\providecommand \selectlanguage [0]{\@gobble}%
\providecommand \bibinfo  [0]{\@secondoftwo}%
\providecommand \bibfield  [0]{\@secondoftwo}%
\providecommand \translation [1]{[#1]}%
\providecommand \BibitemOpen [0]{}%
\providecommand \bibitemStop [0]{}%
\providecommand \bibitemNoStop [0]{.\EOS\space}%
\providecommand \EOS [0]{\spacefactor3000\relax}%
\providecommand \BibitemShut  [1]{\csname bibitem#1\endcsname}%
\let\auto@bib@innerbib\@empty
\bibitem [{\citenamefont {Schomerus}(2013)}]{171007-1}%
  \BibitemOpen
  \bibfield  {author} {\bibinfo {author} {\bibfnamefont {H.}~\bibnamefont
  {Schomerus}},\ }\href {\doibase 10.1364/OL.38.001912} {\bibfield  {journal}
  {\bibinfo  {journal} {Opt. Lett.}\ }\textbf {\bibinfo {volume} {38}},\
  \bibinfo {pages} {1912} (\bibinfo {year} {2013})}\BibitemShut {NoStop}%
\bibitem [{\citenamefont {Su}\ \emph {et~al.}(1979)\citenamefont {Su},
  \citenamefont {Schrieffer},\ and\ \citenamefont {Heeger}}]{130704-1}%
  \BibitemOpen
  \bibfield  {author} {\bibinfo {author} {\bibfnamefont {W.~P.}\ \bibnamefont
  {Su}}, \bibinfo {author} {\bibfnamefont {J.~R.}\ \bibnamefont {Schrieffer}},
  \ and\ \bibinfo {author} {\bibfnamefont {A.~J.}\ \bibnamefont {Heeger}},\
  }\href {\doibase 10.1103/PhysRevLett.42.1698} {\bibfield  {journal} {\bibinfo
   {journal} {Phys. Rev. Lett.}\ }\textbf {\bibinfo {volume} {42}},\ \bibinfo
  {eid} {130704-1} (\bibinfo {year} {1979})}\BibitemShut {NoStop}%
\bibitem [{\citenamefont {Bernevig}\ and\ \citenamefont
  {Hughes}(2013)}]{111222-1}%
  \BibitemOpen
  \bibfield  {author} {\bibinfo {author} {\bibfnamefont {B.}~\bibnamefont
  {Bernevig}}\ and\ \bibinfo {author} {\bibfnamefont {T.}~\bibnamefont
  {Hughes}},\ }\href {https://books.google.com.hk/books?id=\_7r\_UqFN0IEC}
  {\emph {\bibinfo {title} {Topological Insulators and Topological
  Superconductors}}}\ (\bibinfo  {publisher} {Princeton University Press},\
  \bibinfo {year} {2013})\BibitemShut {NoStop}%
\bibitem [{\citenamefont {Poli}\ \emph {et~al.}(2015)\citenamefont {Poli},
  \citenamefont {Bellec}, \citenamefont {Kuhl}, \citenamefont {Mortessagne},\
  and\ \citenamefont {Schomerus}}]{171212-2}%
  \BibitemOpen
  \bibfield  {author} {\bibinfo {author} {\bibfnamefont {C.}~\bibnamefont
  {Poli}}, \bibinfo {author} {\bibfnamefont {M.}~\bibnamefont {Bellec}},
  \bibinfo {author} {\bibfnamefont {U.}~\bibnamefont {Kuhl}}, \bibinfo {author}
  {\bibfnamefont {F.}~\bibnamefont {Mortessagne}}, \ and\ \bibinfo {author}
  {\bibfnamefont {H.}~\bibnamefont {Schomerus}},\ }\href
  {http://dx.doi.org/10.1038/ncomms7710} {\bibfield  {journal} {\bibinfo
  {journal} {Nature Communications}\ }\textbf {\bibinfo {volume} {6}},\
  \bibinfo {pages} {6710} (\bibinfo {year} {2015})}\BibitemShut {NoStop}%
\bibitem [{\citenamefont {Weimann}\ \emph {et~al.}(2016)\citenamefont
  {Weimann}, \citenamefont {Kremer}, \citenamefont {Plotnik}, \citenamefont
  {Lumer}, \citenamefont {Nolte}, \citenamefont {Makris}, \citenamefont
  {Segev}, \citenamefont {Rechtsman},\ and\ \citenamefont
  {Szameit}}]{171207-7}%
  \BibitemOpen
  \bibfield  {author} {\bibinfo {author} {\bibfnamefont {S.}~\bibnamefont
  {Weimann}}, \bibinfo {author} {\bibfnamefont {M.}~\bibnamefont {Kremer}},
  \bibinfo {author} {\bibfnamefont {Y.}~\bibnamefont {Plotnik}}, \bibinfo
  {author} {\bibfnamefont {Y.}~\bibnamefont {Lumer}}, \bibinfo {author}
  {\bibfnamefont {S.}~\bibnamefont {Nolte}}, \bibinfo {author} {\bibfnamefont
  {K.~G.}\ \bibnamefont {Makris}}, \bibinfo {author} {\bibfnamefont
  {M.}~\bibnamefont {Segev}}, \bibinfo {author} {\bibfnamefont
  {M.}~\bibnamefont {Rechtsman}}, \ and\ \bibinfo {author} {\bibfnamefont
  {A.}~\bibnamefont {Szameit}},\ }\href {http://dx.doi.org/10.1038/nmat4811}
  {\bibfield  {journal} {\bibinfo  {journal} {Nature Materials}\ }\textbf
  {\bibinfo {volume} {16}},\ \bibinfo {pages} {433} (\bibinfo {year}
  {2016})}\BibitemShut {NoStop}%
\bibitem [{\citenamefont {Pan}\ \emph {et~al.}(2018)\citenamefont {Pan},
  \citenamefont {Zhao}, \citenamefont {Miao}, \citenamefont {Longhi},\ and\
  \citenamefont {Feng}}]{180409-2}%
  \BibitemOpen
  \bibfield  {author} {\bibinfo {author} {\bibfnamefont {M.}~\bibnamefont
  {Pan}}, \bibinfo {author} {\bibfnamefont {H.}~\bibnamefont {Zhao}}, \bibinfo
  {author} {\bibfnamefont {P.}~\bibnamefont {Miao}}, \bibinfo {author}
  {\bibfnamefont {S.}~\bibnamefont {Longhi}}, \ and\ \bibinfo {author}
  {\bibfnamefont {L.}~\bibnamefont {Feng}},\ }\href
  {https://www.nature.com/articles/s41467-018-03822-8} {\bibfield  {journal}
  {\bibinfo  {journal} {Nature Communications}\ }\textbf {\bibinfo {volume}
  {9}},\ \bibinfo {pages} {1308} (\bibinfo {year} {2018})}\BibitemShut
  {NoStop}%
\bibitem [{\citenamefont {Yao}\ \emph {et~al.}(2018)\citenamefont {Yao},
  \citenamefont {Li}, \citenamefont {Zheng}, \citenamefont {An}, \citenamefont
  {Ding}, \citenamefont {Lee}, \citenamefont {Zhang},\ and\ \citenamefont
  {Guo}}]{180419-3}%
  \BibitemOpen
  \bibfield  {author} {\bibinfo {author} {\bibfnamefont {R.}~\bibnamefont
  {Yao}}, \bibinfo {author} {\bibfnamefont {H.}~\bibnamefont {Li}}, \bibinfo
  {author} {\bibfnamefont {B.}~\bibnamefont {Zheng}}, \bibinfo {author}
  {\bibfnamefont {S.}~\bibnamefont {An}}, \bibinfo {author} {\bibfnamefont
  {J.}~\bibnamefont {Ding}}, \bibinfo {author} {\bibfnamefont {C.-S.}\
  \bibnamefont {Lee}}, \bibinfo {author} {\bibfnamefont {H.}~\bibnamefont
  {Zhang}}, \ and\ \bibinfo {author} {\bibfnamefont {W.}~\bibnamefont {Guo}},\
  }\href@noop {} {\bibfield  {journal} {\bibinfo  {journal} {arXiv:1804.01587}\
  } (\bibinfo {year} {2018})}\BibitemShut {NoStop}%
\bibitem [{\citenamefont {Zhao}\ \emph {et~al.}(2018)\citenamefont {Zhao},
  \citenamefont {Miao}, \citenamefont {Teimourpour}, \citenamefont {Malzard},
  \citenamefont {El-Ganainy}, \citenamefont {Schomerus},\ and\ \citenamefont
  {Feng}}]{180402-1}%
  \BibitemOpen
  \bibfield  {author} {\bibinfo {author} {\bibfnamefont {H.}~\bibnamefont
  {Zhao}}, \bibinfo {author} {\bibfnamefont {P.}~\bibnamefont {Miao}}, \bibinfo
  {author} {\bibfnamefont {M.~H.}\ \bibnamefont {Teimourpour}}, \bibinfo
  {author} {\bibfnamefont {S.}~\bibnamefont {Malzard}}, \bibinfo {author}
  {\bibfnamefont {R.}~\bibnamefont {El-Ganainy}}, \bibinfo {author}
  {\bibfnamefont {H.}~\bibnamefont {Schomerus}}, \ and\ \bibinfo {author}
  {\bibfnamefont {L.}~\bibnamefont {Feng}},\ }\href
  {https://www.nature.com/articles/s41467-018-03434-2} {\bibfield  {journal}
  {\bibinfo  {journal} {Nature Communications}\ }\textbf {\bibinfo {volume}
  {9}},\ \bibinfo {pages} {981} (\bibinfo {year} {2018})}\BibitemShut {NoStop}%
\bibitem [{\citenamefont {Bender}\ and\ \citenamefont
  {Boettcher}(1998)}]{101224-1}%
  \BibitemOpen
  \bibfield  {author} {\bibinfo {author} {\bibfnamefont {C.~M.}\ \bibnamefont
  {Bender}}\ and\ \bibinfo {author} {\bibfnamefont {S.}~\bibnamefont
  {Boettcher}},\ }\href {\doibase 10.1103/PhysRevLett.80.5243} {\bibfield
  {journal} {\bibinfo  {journal} {Phys. Rev. Lett.}\ }\textbf {\bibinfo
  {volume} {80}},\ \bibinfo {eid} {101224-1} (\bibinfo {year}
  {1998})}\BibitemShut {NoStop}%
\bibitem [{\citenamefont {Moiseyev}(2011)}]{110707-1}%
  \BibitemOpen
  \bibfield  {author} {\bibinfo {author} {\bibfnamefont {N.}~\bibnamefont
  {Moiseyev}},\ }\href@noop {} {\emph {\bibinfo {title} {Non-Hermitian Quantum
  Mechanics}}},\ \bibinfo {edition} {1st}\ ed.\ (\bibinfo  {publisher}
  {CAMBRIDGE UNIVERSITY PRESS},\ \bibinfo {year} {2011})\BibitemShut {NoStop}%
\bibitem [{\citenamefont {Ryu}\ \emph {et~al.}(2010)\citenamefont {Ryu},
  \citenamefont {Schnyder}, \citenamefont {Furusaki},\ and\ \citenamefont
  {Ludwig}}]{120316-1}%
  \BibitemOpen
  \bibfield  {author} {\bibinfo {author} {\bibfnamefont {S.}~\bibnamefont
  {Ryu}}, \bibinfo {author} {\bibfnamefont {A.~P.}\ \bibnamefont {Schnyder}},
  \bibinfo {author} {\bibfnamefont {A.}~\bibnamefont {Furusaki}}, \ and\
  \bibinfo {author} {\bibfnamefont {A.~W.~W.}\ \bibnamefont {Ludwig}},\ }\href
  {http://stacks.iop.org/1367-2630/12/i=6/a=065010} {\bibfield  {journal}
  {\bibinfo  {journal} {New Journal of Physics}\ }\textbf {\bibinfo {volume}
  {12}},\ \bibinfo {eid} {120316-1} (\bibinfo {year} {2010})}\BibitemShut
  {NoStop}%
\bibitem [{\citenamefont {Zak}(1989)}]{130107-1}%
  \BibitemOpen
  \bibfield  {author} {\bibinfo {author} {\bibfnamefont {J.}~\bibnamefont
  {Zak}},\ }\href {\doibase 10.1103/PhysRevLett.62.2747} {\bibfield  {journal}
  {\bibinfo  {journal} {Phys. Rev. Lett.}\ }\textbf {\bibinfo {volume} {62}},\
  \bibinfo {eid} {130107-1} (\bibinfo {year} {1989})}\BibitemShut {NoStop}%
\bibitem [{\citenamefont {Harari}\ \emph {et~al.}(2018)\citenamefont {Harari},
  \citenamefont {Bandres}, \citenamefont {Lumer}, \citenamefont {Rechtsman},
  \citenamefont {Chong}, \citenamefont {M.}, \citenamefont {Christodoulides},\
  and\ \citenamefont {Segev}}]{180720-9}%
  \BibitemOpen
  \bibfield  {author} {\bibinfo {author} {\bibfnamefont {G.}~\bibnamefont
  {Harari}}, \bibinfo {author} {\bibfnamefont {M.~A.}\ \bibnamefont {Bandres}},
  \bibinfo {author} {\bibfnamefont {Y.}~\bibnamefont {Lumer}}, \bibinfo
  {author} {\bibfnamefont {M.~C.}\ \bibnamefont {Rechtsman}}, \bibinfo {author}
  {\bibfnamefont {Y.~D.}\ \bibnamefont {Chong}}, \bibinfo {author}
  {\bibfnamefont {K.}~\bibnamefont {M.}}, \bibinfo {author} {\bibfnamefont
  {D.~N.}\ \bibnamefont {Christodoulides}}, \ and\ \bibinfo {author}
  {\bibfnamefont {M.}~\bibnamefont {Segev}},\ }\href@noop {} {\bibfield
  {journal} {\bibinfo  {journal} {Science}\ }\textbf {\bibinfo {volume}
  {359}},\ \bibinfo {pages} {eaar4003} (\bibinfo {year} {2018})}\BibitemShut
  {NoStop}%
\bibitem [{\citenamefont {Lee}(2016)}]{171012-1}%
  \BibitemOpen
  \bibfield  {author} {\bibinfo {author} {\bibfnamefont {T.~E.}\ \bibnamefont
  {Lee}},\ }\href {https://link.aps.org/doi/10.1103/PhysRevLett.116.133903}
  {\bibfield  {journal} {\bibinfo  {journal} {Phys. Rev. Lett.}\ }\textbf
  {\bibinfo {volume} {116}},\ \bibinfo {pages} {133903} (\bibinfo {year}
  {2016})}\BibitemShut {NoStop}%
\bibitem [{\citenamefont {Leykam}\ \emph
  {et~al.}(2017{\natexlab{a}})\citenamefont {Leykam}, \citenamefont {Bliokh},
  \citenamefont {Huang}, \citenamefont {Chong},\ and\ \citenamefont
  {Nori}}]{180720-8}%
  \BibitemOpen
  \bibfield  {author} {\bibinfo {author} {\bibfnamefont {D.}~\bibnamefont
  {Leykam}}, \bibinfo {author} {\bibfnamefont {K.~Y.}\ \bibnamefont {Bliokh}},
  \bibinfo {author} {\bibfnamefont {C.}~\bibnamefont {Huang}}, \bibinfo
  {author} {\bibfnamefont {Y.}~\bibnamefont {Chong}}, \ and\ \bibinfo {author}
  {\bibfnamefont {F.}~\bibnamefont {Nori}},\ }\href@noop {} {\bibfield
  {journal} {\bibinfo  {journal} {Phys. Rev. Lett.}\ }\textbf {\bibinfo
  {volume} {118}},\ \bibinfo {pages} {040401} (\bibinfo {year}
  {2017}{\natexlab{a}})}\BibitemShut {NoStop}%
\bibitem [{\citenamefont {Shen}\ \emph {et~al.}(2018)\citenamefont {Shen},
  \citenamefont {Zhen},\ and\ \citenamefont {Fu}}]{180720-7}%
  \BibitemOpen
  \bibfield  {author} {\bibinfo {author} {\bibfnamefont {H.}~\bibnamefont
  {Shen}}, \bibinfo {author} {\bibfnamefont {B.}~\bibnamefont {Zhen}}, \ and\
  \bibinfo {author} {\bibfnamefont {L.}~\bibnamefont {Fu}},\ }\href {\doibase
  10.1103/PhysRevLett.120.146402} {\bibfield  {journal} {\bibinfo  {journal}
  {Phys. Rev. Lett.}\ }\textbf {\bibinfo {volume} {120}},\ \bibinfo {pages}
  {146402} (\bibinfo {year} {2018})}\BibitemShut {NoStop}%
\bibitem [{\citenamefont {Yin}\ \emph {et~al.}(2018)\citenamefont {Yin},
  \citenamefont {Jiang}, \citenamefont {Li}, \citenamefont {L\"u},\ and\
  \citenamefont {Chen}}]{180720-3}%
  \BibitemOpen
  \bibfield  {author} {\bibinfo {author} {\bibfnamefont {C.}~\bibnamefont
  {Yin}}, \bibinfo {author} {\bibfnamefont {H.}~\bibnamefont {Jiang}}, \bibinfo
  {author} {\bibfnamefont {L.}~\bibnamefont {Li}}, \bibinfo {author}
  {\bibfnamefont {R.}~\bibnamefont {L\"u}}, \ and\ \bibinfo {author}
  {\bibfnamefont {S.}~\bibnamefont {Chen}},\ }\href {\doibase
  10.1103/PhysRevA.97.052115} {\bibfield  {journal} {\bibinfo  {journal} {Phys.
  Rev. A}\ }\textbf {\bibinfo {volume} {97}},\ \bibinfo {pages} {052115}
  (\bibinfo {year} {2018})}\BibitemShut {NoStop}%
\bibitem [{\citenamefont {Kunst}\ \emph {et~al.}(2018)\citenamefont {Kunst},
  \citenamefont {Edvardsson}, \citenamefont {Budich},\ and\ \citenamefont
  {Bergholtz}}]{180720-4}%
  \BibitemOpen
  \bibfield  {author} {\bibinfo {author} {\bibfnamefont {F.~K.}\ \bibnamefont
  {Kunst}}, \bibinfo {author} {\bibfnamefont {E.}~\bibnamefont {Edvardsson}},
  \bibinfo {author} {\bibfnamefont {J.~C.}\ \bibnamefont {Budich}}, \ and\
  \bibinfo {author} {\bibfnamefont {E.~J.}\ \bibnamefont {Bergholtz}},\ }\href
  {\doibase 10.1103/PhysRevLett.121.026808} {\bibfield  {journal} {\bibinfo
  {journal} {Phys. Rev. Lett.}\ }\textbf {\bibinfo {volume} {121}},\ \bibinfo
  {pages} {026808} (\bibinfo {year} {2018})}\BibitemShut {NoStop}%
\bibitem [{\citenamefont {Gong}\ \emph {et~al.}(2018)\citenamefont {Gong},
  \citenamefont {Ashida}, \citenamefont {Kawabata}, \citenamefont {Takasan},
  \citenamefont {Higashikawa},\ and\ \citenamefont {Ueda}}]{180720-5}%
  \BibitemOpen
  \bibfield  {author} {\bibinfo {author} {\bibfnamefont {Z.}~\bibnamefont
  {Gong}}, \bibinfo {author} {\bibfnamefont {Y.}~\bibnamefont {Ashida}},
  \bibinfo {author} {\bibfnamefont {K.}~\bibnamefont {Kawabata}}, \bibinfo
  {author} {\bibfnamefont {K.}~\bibnamefont {Takasan}}, \bibinfo {author}
  {\bibfnamefont {S.}~\bibnamefont {Higashikawa}}, \ and\ \bibinfo {author}
  {\bibfnamefont {M.}~\bibnamefont {Ueda}},\ }\href@noop {} {\bibfield
  {journal} {\bibinfo  {journal} {arXiv:1802.07964}\ } (\bibinfo {year}
  {2018})}\BibitemShut {NoStop}%
\bibitem [{\citenamefont {Xiong}(2018)}]{180720-6}%
  \BibitemOpen
  \bibfield  {author} {\bibinfo {author} {\bibfnamefont {Y.}~\bibnamefont
  {Xiong}},\ }\href {http://stacks.iop.org/2399-6528/2/i=3/a=035043} {\bibfield
   {journal} {\bibinfo  {journal} {Journal of Physics Communications}\ }\textbf
  {\bibinfo {volume} {2}},\ \bibinfo {pages} {035043} (\bibinfo {year}
  {2018})}\BibitemShut {NoStop}%
\bibitem [{\citenamefont {Zhu}\ \emph {et~al.}(2014)\citenamefont {Zhu},
  \citenamefont {L\"u},\ and\ \citenamefont {Chen}}]{180720-2}%
  \BibitemOpen
  \bibfield  {author} {\bibinfo {author} {\bibfnamefont {B.}~\bibnamefont
  {Zhu}}, \bibinfo {author} {\bibfnamefont {R.}~\bibnamefont {L\"u}}, \ and\
  \bibinfo {author} {\bibfnamefont {S.}~\bibnamefont {Chen}},\ }\href {\doibase
  10.1103/PhysRevA.89.062102} {\bibfield  {journal} {\bibinfo  {journal} {Phys.
  Rev. A}\ }\textbf {\bibinfo {volume} {89}},\ \bibinfo {pages} {062102}
  (\bibinfo {year} {2014})}\BibitemShut {NoStop}%
\bibitem [{\citenamefont {Lu}\ \emph {et~al.}(2014)\citenamefont {Lu},
  \citenamefont {Joannopoulos},\ and\ \citenamefont
  {Solja{\v{c}}i{\'c}}}]{tpreview2014}%
  \BibitemOpen
  \bibfield  {author} {\bibinfo {author} {\bibfnamefont {L.}~\bibnamefont
  {Lu}}, \bibinfo {author} {\bibfnamefont {J.~D.}\ \bibnamefont
  {Joannopoulos}}, \ and\ \bibinfo {author} {\bibfnamefont {M.}~\bibnamefont
  {Solja{\v{c}}i{\'c}}},\ }\href@noop {} {\bibfield  {journal} {\bibinfo
  {journal} {Nat. Photon.}\ }\textbf {\bibinfo {volume} {8}},\ \bibinfo {pages}
  {821} (\bibinfo {year} {2014})}\BibitemShut {NoStop}%
\bibitem [{\citenamefont {Khanikaev}\ and\ \citenamefont
  {Shvets}(2017)}]{tpreview2017}%
  \BibitemOpen
  \bibfield  {author} {\bibinfo {author} {\bibfnamefont {A.~B.}\ \bibnamefont
  {Khanikaev}}\ and\ \bibinfo {author} {\bibfnamefont {G.}~\bibnamefont
  {Shvets}},\ }\href@noop {} {\bibfield  {journal} {\bibinfo  {journal} {Nat.
  Photon.}\ }\textbf {\bibinfo {volume} {11}},\ \bibinfo {pages} {763}
  (\bibinfo {year} {2017})}\BibitemShut {NoStop}%
\bibitem [{\citenamefont {Ozawa}\ \emph {et~al.}()\citenamefont {Ozawa},
  \citenamefont {Price}, \citenamefont {Amo}, \citenamefont {Goldman},
  \citenamefont {Hafezi}, \citenamefont {Lu}, \citenamefont {Rechtsman},
  \citenamefont {Schuster}, \citenamefont {Simon}, \citenamefont {Zilberberg},\
  and\ \citenamefont {Carusotto}}]{tpreview2018}%
  \BibitemOpen
  \bibfield  {author} {\bibinfo {author} {\bibfnamefont {T.}~\bibnamefont
  {Ozawa}}, \bibinfo {author} {\bibfnamefont {H.~M.}\ \bibnamefont {Price}},
  \bibinfo {author} {\bibfnamefont {A.}~\bibnamefont {Amo}}, \bibinfo {author}
  {\bibfnamefont {N.}~\bibnamefont {Goldman}}, \bibinfo {author} {\bibfnamefont
  {M.}~\bibnamefont {Hafezi}}, \bibinfo {author} {\bibfnamefont
  {L.}~\bibnamefont {Lu}}, \bibinfo {author} {\bibfnamefont {M.~C.}\
  \bibnamefont {Rechtsman}}, \bibinfo {author} {\bibfnamefont {D.}~\bibnamefont
  {Schuster}}, \bibinfo {author} {\bibfnamefont {J.}~\bibnamefont {Simon}},
  \bibinfo {author} {\bibfnamefont {O.}~\bibnamefont {Zilberberg}}, \ and\
  \bibinfo {author} {\bibfnamefont {I.}~\bibnamefont {Carusotto}},\ }\href@noop
  {} {\enquote {\bibinfo {title} {Topological photonics},}\ }\Eprint
  {http://arxiv.org/abs/arXiv:1802.04173} {arXiv:1802.04173} \BibitemShut
  {NoStop}%
\bibitem [{\citenamefont {Hu}\ \emph {et~al.}(2017)\citenamefont {Hu},
  \citenamefont {Wang}, \citenamefont {Shum},\ and\ \citenamefont
  {Chong}}]{hailong1}%
  \BibitemOpen
  \bibfield  {author} {\bibinfo {author} {\bibfnamefont {W.}~\bibnamefont
  {Hu}}, \bibinfo {author} {\bibfnamefont {H.}~\bibnamefont {Wang}}, \bibinfo
  {author} {\bibfnamefont {P.~P.}\ \bibnamefont {Shum}}, \ and\ \bibinfo
  {author} {\bibfnamefont {Y.~D.}\ \bibnamefont {Chong}},\ }\href {\doibase
  10.1103/PhysRevB.95.184306} {\bibfield  {journal} {\bibinfo  {journal} {Phys.
  Rev. B}\ }\textbf {\bibinfo {volume} {95}},\ \bibinfo {pages} {184306}
  (\bibinfo {year} {2017})}\BibitemShut {NoStop}%
\bibitem [{\citenamefont {Wang}\ \emph {et~al.}(2018)\citenamefont {Wang},
  \citenamefont {Lang},\ and\ \citenamefont {Chong}}]{hailong2}%
  \BibitemOpen
  \bibfield  {author} {\bibinfo {author} {\bibfnamefont {H.}~\bibnamefont
  {Wang}}, \bibinfo {author} {\bibfnamefont {L.-J.}\ \bibnamefont {Lang}}, \
  and\ \bibinfo {author} {\bibfnamefont {Y.~D.}\ \bibnamefont {Chong}},\ }\href
  {\doibase 10.1103/PhysRevA.98.012119} {\bibfield  {journal} {\bibinfo
  {journal} {Phys. Rev. A}\ }\textbf {\bibinfo {volume} {98}},\ \bibinfo
  {pages} {012119} (\bibinfo {year} {2018})}\BibitemShut {NoStop}%
\bibitem [{\citenamefont {Makris}\ \emph {et~al.}(2008)\citenamefont {Makris},
  \citenamefont {El-Ganainy}, \citenamefont {Christodoulides},\ and\
  \citenamefont {Musslimani}}]{180501-3}%
  \BibitemOpen
  \bibfield  {author} {\bibinfo {author} {\bibfnamefont {K.~G.}\ \bibnamefont
  {Makris}}, \bibinfo {author} {\bibfnamefont {R.}~\bibnamefont {El-Ganainy}},
  \bibinfo {author} {\bibfnamefont {D.~N.}\ \bibnamefont {Christodoulides}}, \
  and\ \bibinfo {author} {\bibfnamefont {Z.~H.}\ \bibnamefont {Musslimani}},\
  }\href {\doibase 10.1103/PhysRevLett.100.103904} {\bibfield  {journal}
  {\bibinfo  {journal} {Phys. Rev. Lett.}\ }\textbf {\bibinfo {volume} {100}},\
  \bibinfo {pages} {103904} (\bibinfo {year} {2008})}\BibitemShut {NoStop}%
\bibitem [{\citenamefont {R\"{u}ter}\ \emph {et~al.}(2010)\citenamefont
  {R\"{u}ter}, \citenamefont {Makris}, \citenamefont {El-Ganainy},
  \citenamefont {Christodoulides}, \citenamefont {Segev},\ and\ \citenamefont
  {Kip}}]{180501-1}%
  \BibitemOpen
  \bibfield  {author} {\bibinfo {author} {\bibfnamefont {C.~E.}\ \bibnamefont
  {R\"{u}ter}}, \bibinfo {author} {\bibfnamefont {K.~G.}\ \bibnamefont
  {Makris}}, \bibinfo {author} {\bibfnamefont {R.}~\bibnamefont {El-Ganainy}},
  \bibinfo {author} {\bibfnamefont {D.~N.}\ \bibnamefont {Christodoulides}},
  \bibinfo {author} {\bibfnamefont {M.}~\bibnamefont {Segev}}, \ and\ \bibinfo
  {author} {\bibfnamefont {D.}~\bibnamefont {Kip}},\ }\href
  {http://dx.doi.org/10.1038/nphys1515} {\bibfield  {journal} {\bibinfo
  {journal} {Nature Physics}\ }\textbf {\bibinfo {volume} {6}},\ \bibinfo
  {pages} {192} (\bibinfo {year} {2010})}\BibitemShut {NoStop}%
\bibitem [{\citenamefont {Liang}\ and\ \citenamefont
  {Chong}(2013)}]{liang2013}%
  \BibitemOpen
  \bibfield  {author} {\bibinfo {author} {\bibfnamefont {G.~Q.}\ \bibnamefont
  {Liang}}\ and\ \bibinfo {author} {\bibfnamefont {Y.~D.}\ \bibnamefont
  {Chong}},\ }\href@noop {} {\bibfield  {journal} {\bibinfo  {journal} {Phys.
  Rev. Lett.}\ }\textbf {\bibinfo {volume} {110}},\ \bibinfo {pages} {203904}
  (\bibinfo {year} {2013})}\BibitemShut {NoStop}%
\bibitem [{\citenamefont {Peano}\ \emph {et~al.}(2016)\citenamefont {Peano},
  \citenamefont {Houde}, \citenamefont {Marquardt},\ and\ \citenamefont
  {Clerk}}]{Peano2016}%
  \BibitemOpen
  \bibfield  {author} {\bibinfo {author} {\bibfnamefont {V.}~\bibnamefont
  {Peano}}, \bibinfo {author} {\bibfnamefont {M.}~\bibnamefont {Houde}},
  \bibinfo {author} {\bibfnamefont {F.}~\bibnamefont {Marquardt}}, \ and\
  \bibinfo {author} {\bibfnamefont {A.~A.}\ \bibnamefont {Clerk}},\ }\href@noop
  {} {\bibfield  {journal} {\bibinfo  {journal} {Phys. Rev. X}\ }\textbf
  {\bibinfo {volume} {6}},\ \bibinfo {pages} {041026} (\bibinfo {year}
  {2016})}\BibitemShut {NoStop}%
\bibitem [{\citenamefont {St-Jean}\ \emph {et~al.}(2017)\citenamefont
  {St-Jean}, \citenamefont {Goblot}, \citenamefont {Galopin}, \citenamefont
  {Lemaitre}, \citenamefont {Ozawa}, \citenamefont {Le~Gratiet}, \citenamefont
  {Sagnes}, \citenamefont {Bloch},\ and\ \citenamefont {Amo}}]{180719-1}%
  \BibitemOpen
  \bibfield  {author} {\bibinfo {author} {\bibfnamefont {P.}~\bibnamefont
  {St-Jean}}, \bibinfo {author} {\bibfnamefont {V.}~\bibnamefont {Goblot}},
  \bibinfo {author} {\bibfnamefont {E.}~\bibnamefont {Galopin}}, \bibinfo
  {author} {\bibfnamefont {A.}~\bibnamefont {Lemaitre}}, \bibinfo {author}
  {\bibfnamefont {T.}~\bibnamefont {Ozawa}}, \bibinfo {author} {\bibfnamefont
  {L.}~\bibnamefont {Le~Gratiet}}, \bibinfo {author} {\bibfnamefont
  {I.}~\bibnamefont {Sagnes}}, \bibinfo {author} {\bibfnamefont
  {J.}~\bibnamefont {Bloch}}, \ and\ \bibinfo {author} {\bibfnamefont
  {A.}~\bibnamefont {Amo}},\ }\href {https://doi.org/10.1038/s41566-017-0006-2}
  {\bibfield  {journal} {\bibinfo  {journal} {Nature Photonics}\ }\textbf
  {\bibinfo {volume} {11}},\ \bibinfo {pages} {651} (\bibinfo {year}
  {2017})}\BibitemShut {NoStop}%
\bibitem [{\citenamefont {Malzard}\ and\ \citenamefont
  {Schomerus}(2018{\natexlab{a}})}]{180719-2}%
  \BibitemOpen
  \bibfield  {author} {\bibinfo {author} {\bibfnamefont {S.}~\bibnamefont
  {Malzard}}\ and\ \bibinfo {author} {\bibfnamefont {H.}~\bibnamefont
  {Schomerus}},\ }\href {http://stacks.iop.org/1367-2630/20/i=6/a=063044}
  {\bibfield  {journal} {\bibinfo  {journal} {New Journal of Physics}\ }\textbf
  {\bibinfo {volume} {20}},\ \bibinfo {pages} {063044} (\bibinfo {year}
  {2018}{\natexlab{a}})}\BibitemShut {NoStop}%
\bibitem [{\citenamefont {Bandres}\ \emph {et~al.}(2018)\citenamefont
  {Bandres}, \citenamefont {Wittek}, \citenamefont {Harari}, \citenamefont
  {Parto}, \citenamefont {Ren}, \citenamefont {Segev}, \citenamefont
  {Christodoulides},\ and\ \citenamefont {Khajavikhan}}]{180720-10}%
  \BibitemOpen
  \bibfield  {author} {\bibinfo {author} {\bibfnamefont {M.~A.}\ \bibnamefont
  {Bandres}}, \bibinfo {author} {\bibfnamefont {S.}~\bibnamefont {Wittek}},
  \bibinfo {author} {\bibfnamefont {G.}~\bibnamefont {Harari}}, \bibinfo
  {author} {\bibfnamefont {M.}~\bibnamefont {Parto}}, \bibinfo {author}
  {\bibfnamefont {J.}~\bibnamefont {Ren}}, \bibinfo {author} {\bibfnamefont
  {M.}~\bibnamefont {Segev}}, \bibinfo {author} {\bibfnamefont {D.~N.}\
  \bibnamefont {Christodoulides}}, \ and\ \bibinfo {author} {\bibfnamefont
  {M.}~\bibnamefont {Khajavikhan}},\ }\href@noop {} {\bibfield  {journal}
  {\bibinfo  {journal} {Science}\ }\textbf {\bibinfo {volume} {359}},\ \bibinfo
  {pages} {eaar4005} (\bibinfo {year} {2018})}\BibitemShut {NoStop}%
\bibitem [{\citenamefont {Guo}\ \emph {et~al.}(2009)\citenamefont {Guo},
  \citenamefont {Salamo}, \citenamefont {Duchesne}, \citenamefont {Morandotti},
  \citenamefont {Volatier-Ravat}, \citenamefont {Aimez}, \citenamefont
  {Siviloglou},\ and\ \citenamefont {Christodoulides}}]{180501-2}%
  \BibitemOpen
  \bibfield  {author} {\bibinfo {author} {\bibfnamefont {A.}~\bibnamefont
  {Guo}}, \bibinfo {author} {\bibfnamefont {G.~J.}\ \bibnamefont {Salamo}},
  \bibinfo {author} {\bibfnamefont {D.}~\bibnamefont {Duchesne}}, \bibinfo
  {author} {\bibfnamefont {R.}~\bibnamefont {Morandotti}}, \bibinfo {author}
  {\bibfnamefont {M.}~\bibnamefont {Volatier-Ravat}}, \bibinfo {author}
  {\bibfnamefont {V.}~\bibnamefont {Aimez}}, \bibinfo {author} {\bibfnamefont
  {G.~A.}\ \bibnamefont {Siviloglou}}, \ and\ \bibinfo {author} {\bibfnamefont
  {D.~N.}\ \bibnamefont {Christodoulides}},\ }\href {\doibase
  10.1103/PhysRevLett.103.093902} {\bibfield  {journal} {\bibinfo  {journal}
  {Phys. Rev. Lett.}\ }\textbf {\bibinfo {volume} {103}},\ \bibinfo {pages}
  {093902} (\bibinfo {year} {2009})}\BibitemShut {NoStop}%
\bibitem [{\citenamefont {Malzard}\ \emph {et~al.}(2015)\citenamefont
  {Malzard}, \citenamefont {Poli},\ and\ \citenamefont {Schomerus}}]{171009-1}%
  \BibitemOpen
  \bibfield  {author} {\bibinfo {author} {\bibfnamefont {S.}~\bibnamefont
  {Malzard}}, \bibinfo {author} {\bibfnamefont {C.}~\bibnamefont {Poli}}, \
  and\ \bibinfo {author} {\bibfnamefont {H.}~\bibnamefont {Schomerus}},\ }\href
  {\doibase 10.1103/PhysRevLett.115.200402} {\bibfield  {journal} {\bibinfo
  {journal} {Phys. Rev. Lett.}\ }\textbf {\bibinfo {volume} {115}},\ \bibinfo
  {pages} {200402} (\bibinfo {year} {2015})}\BibitemShut {NoStop}%
\bibitem [{\citenamefont {Takata}\ and\ \citenamefont
  {Notomi}(2017)}]{180306-1}%
  \BibitemOpen
  \bibfield  {author} {\bibinfo {author} {\bibfnamefont {K.}~\bibnamefont
  {Takata}}\ and\ \bibinfo {author} {\bibfnamefont {M.}~\bibnamefont
  {Notomi}},\ }\href@noop {} {\bibfield  {journal} {\bibinfo  {journal}
  {arXiv:1710.09169v1}\ } (\bibinfo {year} {2017})}\BibitemShut {NoStop}%
\bibitem [{\citenamefont {Qi}\ \emph {et~al.}(2018)\citenamefont {Qi},
  \citenamefont {Zhang},\ and\ \citenamefont {Ge}}]{180322-1}%
  \BibitemOpen
  \bibfield  {author} {\bibinfo {author} {\bibfnamefont {B.}~\bibnamefont
  {Qi}}, \bibinfo {author} {\bibfnamefont {L.}~\bibnamefont {Zhang}}, \ and\
  \bibinfo {author} {\bibfnamefont {L.}~\bibnamefont {Ge}},\ }\href {\doibase
  10.1103/PhysRevLett.120.093901} {\bibfield  {journal} {\bibinfo  {journal}
  {Phys. Rev. Lett.}\ }\textbf {\bibinfo {volume} {120}},\ \bibinfo {pages}
  {093901} (\bibinfo {year} {2018})}\BibitemShut {NoStop}%
\bibitem [{\citenamefont {Malzard}\ and\ \citenamefont
  {Schomerus}(2018{\natexlab{b}})}]{180613-1}%
  \BibitemOpen
  \bibfield  {author} {\bibinfo {author} {\bibfnamefont {S.}~\bibnamefont
  {Malzard}}\ and\ \bibinfo {author} {\bibfnamefont {H.}~\bibnamefont
  {Schomerus}},\ }\href@noop {} {\bibfield  {journal} {\bibinfo  {journal}
  {arXiv:1805.08161}\ } (\bibinfo {year} {2018}{\natexlab{b}})}\BibitemShut
  {NoStop}%
\bibitem [{\citenamefont {Leykam}\ \emph
  {et~al.}(2017{\natexlab{b}})\citenamefont {Leykam}, \citenamefont {Flach},\
  and\ \citenamefont {Chong}}]{180720-1}%
  \BibitemOpen
  \bibfield  {author} {\bibinfo {author} {\bibfnamefont {D.}~\bibnamefont
  {Leykam}}, \bibinfo {author} {\bibfnamefont {S.}~\bibnamefont {Flach}}, \
  and\ \bibinfo {author} {\bibfnamefont {Y.~D.}\ \bibnamefont {Chong}},\ }\href
  {\doibase 10.1103/PhysRevB.96.064305} {\bibfield  {journal} {\bibinfo
  {journal} {Phys. Rev. B}\ }\textbf {\bibinfo {volume} {96}},\ \bibinfo
  {pages} {064305} (\bibinfo {year} {2017}{\natexlab{b}})}\BibitemShut
  {NoStop}%
\bibitem [{\citenamefont {Yuce}(2018)}]{180614-1}%
  \BibitemOpen
  \bibfield  {author} {\bibinfo {author} {\bibfnamefont {C.}~\bibnamefont
  {Yuce}},\ }\href {\doibase 10.1103/PhysRevA.97.042118} {\bibfield  {journal}
  {\bibinfo  {journal} {Phys. Rev. A}\ }\textbf {\bibinfo {volume} {97}},\
  \bibinfo {pages} {042118} (\bibinfo {year} {2018})}\BibitemShut {NoStop}%
\bibitem [{\citenamefont {von Neuman}\ and\ \citenamefont
  {Wigner}(1929)}]{180703-2}%
  \BibitemOpen
  \bibfield  {author} {\bibinfo {author} {\bibfnamefont {J.}~\bibnamefont {von
  Neuman}}\ and\ \bibinfo {author} {\bibfnamefont {E.}~\bibnamefont {Wigner}},\
  }\href@noop {} {\bibfield  {journal} {\bibinfo  {journal} {Physikalische
  Zeitschrift}\ }\textbf {\bibinfo {volume} {30}},\ \bibinfo {pages} {467}
  (\bibinfo {year} {1929})}\BibitemShut {NoStop}%
\bibitem [{\citenamefont {Hsu}\ \emph {et~al.}(2016)\citenamefont {Hsu},
  \citenamefont {Zhen}, \citenamefont {Stone}, \citenamefont {Joannopoulos},\
  and\ \citenamefont {Soljacic}}]{180703-1}%
  \BibitemOpen
  \bibfield  {author} {\bibinfo {author} {\bibfnamefont {C.~W.}\ \bibnamefont
  {Hsu}}, \bibinfo {author} {\bibfnamefont {B.}~\bibnamefont {Zhen}}, \bibinfo
  {author} {\bibfnamefont {A.~D.}\ \bibnamefont {Stone}}, \bibinfo {author}
  {\bibfnamefont {J.~D.}\ \bibnamefont {Joannopoulos}}, \ and\ \bibinfo
  {author} {\bibfnamefont {M.}~\bibnamefont {Soljacic}},\ }\href
  {http://dx.doi.org/10.1038/natrevmats.2016.48} {\bibfield  {journal}
  {\bibinfo  {journal} {Nature Reviews Materials}\ }\textbf {\bibinfo {volume}
  {1}},\ \bibinfo {pages} {16048} (\bibinfo {year} {2016})}\BibitemShut
  {NoStop}%
\bibitem [{\citenamefont {Regensburger}\ \emph {et~al.}(2013)\citenamefont
  {Regensburger}, \citenamefont {Miri}, \citenamefont {Bersch}, \citenamefont
  {N\"ager}, \citenamefont {Onishchukov}, \citenamefont {Christodoulides},\
  and\ \citenamefont {Peschel}}]{180703-4}%
  \BibitemOpen
  \bibfield  {author} {\bibinfo {author} {\bibfnamefont {A.}~\bibnamefont
  {Regensburger}}, \bibinfo {author} {\bibfnamefont {M.-A.}\ \bibnamefont
  {Miri}}, \bibinfo {author} {\bibfnamefont {C.}~\bibnamefont {Bersch}},
  \bibinfo {author} {\bibfnamefont {J.}~\bibnamefont {N\"ager}}, \bibinfo
  {author} {\bibfnamefont {G.}~\bibnamefont {Onishchukov}}, \bibinfo {author}
  {\bibfnamefont {D.~N.}\ \bibnamefont {Christodoulides}}, \ and\ \bibinfo
  {author} {\bibfnamefont {U.}~\bibnamefont {Peschel}},\ }\href {\doibase
  10.1103/PhysRevLett.110.223902} {\bibfield  {journal} {\bibinfo  {journal}
  {Phys. Rev. Lett.}\ }\textbf {\bibinfo {volume} {110}},\ \bibinfo {pages}
  {223902} (\bibinfo {year} {2013})}\BibitemShut {NoStop}%
\bibitem [{\citenamefont {Longhi}(2014)}]{180703-3}%
  \BibitemOpen
  \bibfield  {author} {\bibinfo {author} {\bibfnamefont {S.}~\bibnamefont
  {Longhi}},\ }\href {\doibase 10.1364/OL.39.001697} {\bibfield  {journal}
  {\bibinfo  {journal} {Opt. Lett.}\ }\textbf {\bibinfo {volume} {39}},\
  \bibinfo {pages} {1697} (\bibinfo {year} {2014})}\BibitemShut {NoStop}%
\bibitem [{\citenamefont {Kartashov}\ \emph {et~al.}(2018)\citenamefont
  {Kartashov}, \citenamefont {Mili\'{a}n}, \citenamefont {Konotop},\ and\
  \citenamefont {Torner}}]{180703-5}%
  \BibitemOpen
  \bibfield  {author} {\bibinfo {author} {\bibfnamefont {Y.~V.}\ \bibnamefont
  {Kartashov}}, \bibinfo {author} {\bibfnamefont {C.}~\bibnamefont
  {Mili\'{a}n}}, \bibinfo {author} {\bibfnamefont {V.~V.}\ \bibnamefont
  {Konotop}}, \ and\ \bibinfo {author} {\bibfnamefont {L.}~\bibnamefont
  {Torner}},\ }\href {\doibase 10.1364/OL.43.000575} {\bibfield  {journal}
  {\bibinfo  {journal} {Opt. Lett.}\ }\textbf {\bibinfo {volume} {43}},\
  \bibinfo {pages} {575} (\bibinfo {year} {2018})}\BibitemShut {NoStop}%
\bibitem [{\citenamefont {Parto}\ \emph {et~al.}(2018)\citenamefont {Parto},
  \citenamefont {Wittek}, \citenamefont {Hodaei}, \citenamefont {Harari},
  \citenamefont {Bandres}, \citenamefont {Ren}, \citenamefont {Rechtsman},
  \citenamefont {Segev}, \citenamefont {Christodoulides},\ and\ \citenamefont
  {Khajavikhan}}]{180409-1}%
  \BibitemOpen
  \bibfield  {author} {\bibinfo {author} {\bibfnamefont {M.}~\bibnamefont
  {Parto}}, \bibinfo {author} {\bibfnamefont {S.}~\bibnamefont {Wittek}},
  \bibinfo {author} {\bibfnamefont {H.}~\bibnamefont {Hodaei}}, \bibinfo
  {author} {\bibfnamefont {G.}~\bibnamefont {Harari}}, \bibinfo {author}
  {\bibfnamefont {M.~A.}\ \bibnamefont {Bandres}}, \bibinfo {author}
  {\bibfnamefont {J.}~\bibnamefont {Ren}}, \bibinfo {author} {\bibfnamefont
  {M.~C.}\ \bibnamefont {Rechtsman}}, \bibinfo {author} {\bibfnamefont
  {M.}~\bibnamefont {Segev}}, \bibinfo {author} {\bibfnamefont {D.~N.}\
  \bibnamefont {Christodoulides}}, \ and\ \bibinfo {author} {\bibfnamefont
  {M.}~\bibnamefont {Khajavikhan}},\ }\href {\doibase
  10.1103/PhysRevLett.120.113901} {\bibfield  {journal} {\bibinfo  {journal}
  {Phys. Rev. Lett.}\ }\textbf {\bibinfo {volume} {120}},\ \bibinfo {pages}
  {113901} (\bibinfo {year} {2018})}\BibitemShut {NoStop}%
\end{thebibliography}

\end{document}